\documentclass[a4paper,11pt]{article}
\pdfoutput=1 

\usepackage{jheppub} 
\usepackage[T1]{fontenc} 
\usepackage{float}
\usepackage[usenames,dvipsnames]{xcolor}
\usepackage{mathtools}
\usepackage[]{mdframed}
\usepackage{amssymb}

\title{\boldmath Nesting is Not Contracting}

\author{Bart{\l}omiej Czech}
\author{and Sirui Shuai}
\affiliation{Institute for Advanced Study, Tsinghua University, Beijing 100084, China}

\emailAdd{bartlomiej.czech@gmail.com}
\emailAdd{siruishuai@gmail.com}

\abstract{The default way of proving holographic entropy inequalities is the contraction method. It divides Ryu-Takayanagi (RT) surfaces on the `greater than' side of the inequality into segments, then glues the segments into candidate RT surfaces for terms on the `less than' side. Here we discuss how proofs by contraction are constrained and informed by entanglement wedge nesting (EWN)---the property that enlarging a boundary region can only enlarge its entanglement wedge. We propose that: (i) all proofs by contraction necessarily involve candidate RT surfaces, which violate EWN; (ii) violations of EWN in contraction proofs of maximally tight inequalities occur commonly and---where this can be quantified---with maximal density near boundary conditions; (iii) the non-uniqueness of proofs by contraction reflects inequivalent ways of violating EWN. As evidence and illustration, we study the recently discovered infinite families of holographic entropy inequalities, which are associated with tessellations of the torus and the projective plane. We explain the logic, which underlies their proofs by contraction. We find that all salient aspects of the requisite contraction maps are dictated by EWN while all their variable aspects set the scheme for how to violate EWN. We comment on whether the tension between EWN and contraction maps might help in characterizing maximally tight holographic entropy inequalities.}

\begin{document} 
\maketitle
\flushbottom

\section{Introduction and review}
The relationship between the geometry of the spacetime manifold and quantum information has been one of the most intriguing and fruitful ideas in theoretical physics in the last decades. Originated by Bekenstein and Hawking in their study of black hole entropy, it took on a center stage after the formulation of the AdS/CFT correspondence \cite{Maldacena:1997re,Witten:1998qj} and the Ryu-Takayanagi (RT) proposal \cite{Ryu:2006bv, rt2}. Under the assumption of time reversal symmetry, the RT proposal posits that entanglement entropies on the CFT side of the duality are manifested on the bulk (AdS) side as areas of minimal surfaces. Generalizations of the proposal incorporate time dependence \cite{Hubeny:2007xt,Wall:2012uf} and quantum corrections \cite{Engelhardt:2014gca}, but this paper is concerned with the original RT proposal. We assume time reversal symmetry and work at leading order in $G_N$.

Soon after the formulation of the Ryu-Takayanagi proposal it was realized that not all quantum states have entropies, which are viable candidates to describe areas of minimal surfaces. The first necessary condition for an area-like interpretation of entanglement entropies is the monogamy of mutual information \cite{Hayden:2011ag}:
\begin{equation}
S_{AB} + S_{BC} + S_{CA} \geq S_A + S_B + S_C + S_{ABC}
\label{mmi1}
\end{equation}
Examples of states that violate (\ref{mmi1}) include the GHZ state on four parties and the vacuum of the 1+1-dimensional free boson \cite{freeboson}. Indeed, these states cannot be the holographic description of a semiclassical spacetime geometry. 

Since then, many other, more complicated necessary conditions have been discovered \cite{HernandezCuenca:2019wgh, Czech:2022fzb, Hernandez-Cuenca:2023iqh}, including two infinite families \cite{Czech:2023xed, liuyuyubo}. We review them in the beginning of Section~\ref{sec:violation}. Collectively, these inequalities are said to carve the holographic entropy cone because each of them eliminates half of the linear space of potential assignments of entropies to regions (entropy space). For extensions of the holographic entropy cone that incorporate time dependence and quantum effects, see \cite{Czech:2019lps, mattnew, sergioqes}. The holographic entropy cone can also be used to detect non-random tensors in tensor network states. At large bond dimension, random tensor network (RTN) states \cite{Hayden:2016cfa}---states prepared by tensor networks of arbitrary architecture but only using random tensors---satisfy all the holographic inequalities. Therefore, a violation of any holographic entropy inequality in a tensor network state amounts to a detection of non-randomness in at least one tensor.

Thus far, all holographic inequalities other than (\ref{mmi1}) have been proven using a single method called proofs by contraction \cite{Bao:2015bfa}. We review this proof method in Section~\ref{sec:contractreview}. The main goal of this paper is to point out a complex interplay between proofs by contraction and the entanglement wedge nesting theorem, which we review in Section~\ref{sec:ewnreview}. 

This interplay is subtle and ambivalent. On the one hand, it partly undermines the physical motivation for proofs by contraction. On the other hand, it can be very helpful in actually conducting the proofs. In the case of the two infinite families of holographic inequalities \cite{Czech:2023xed}, it reveals a highly non-obvious topological structure, which underlies the form and meaning of the inequalities. We develop these statements in Sections~\ref{sec:violation} and \ref{sec:study}. 

\subsection{Review of proofs by contraction}
\label{sec:contractreview}

We first set the conventions. We write every inequality in the form ${\rm LHS} \geq {\rm RHS}$, where both sides only feature terms with positive coefficients. For example, (\ref{mmi1}) adheres to this convention. Schematically, every inequality takes the form:
\begin{equation}
{\rm LHS} =:\,\, \sum_{i=1}^l \alpha_i S_{X_i} \,\geq\, \sum_{j=1}^r S_{Y_j} \,\,:= {\rm RHS}
\label{generalstr}
\end{equation}
The coefficients $\alpha_i$ are positive. The $X_i$ and $Y_j$ are unions of fundamental regions,\footnote{Fundamental regions are often called monochromatic in the literature.} which in inequality~(\ref{mmi1}) are $A,B,C$, and the implicit purifier $O$. Quantity $l$ counts \emph{distinct} terms on the LHS whereas $r$ counts \emph{all} terms on the RHS, distinct or identical. (We can have $Y_j = Y_{j'}$ but not $X_i = X_{i'}$.) The reason for setting up the notation in this way will be clarified shortly. 

\begin{figure}[t]
    \centering
    \scalebox{1}[1]{\includegraphics[width=0.8\linewidth]{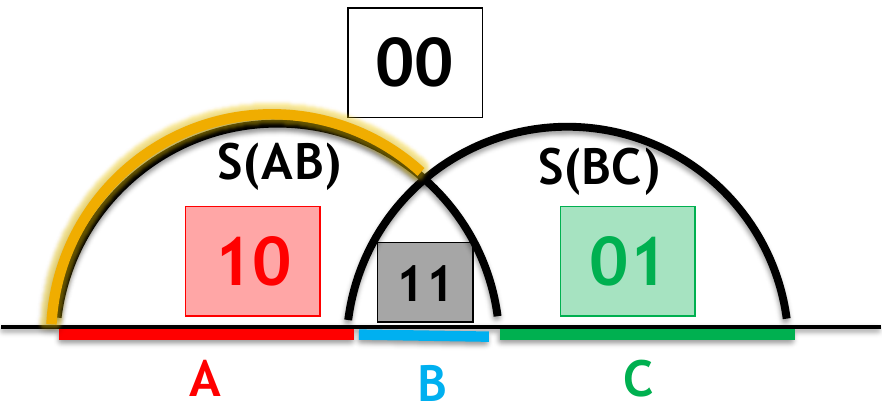}}
    \caption{Regions $W(x)$, which appear in the proof by contraction of $S_{A} + S_B \geq S_{AB}$. This inequality has $l=2$ pieces so there are $2^l = 4$ regions $W(x)$. The highlighted piece of the $S_{AB}$ RT surface is the segment that separates $W(00)$ from $W(10)$; see equations~(\ref{lhseval}) and (\ref{rhseval}).}
    \label{fig:wx}
\end{figure}

Let us associate a binary bit (1 or 0) with every term on the `greater than' side (LHS) of the inequality. We will refer to such an assignment as a {\bf bit string} $x \in \{0,1\}^l$. To be concrete, let us fix one region on the LHS, for example $X_1 = AB$ in inequality~(\ref{mmi1}). We interpret the $AB$ bit in $x$ as dividing the bulk into two halves. If the bit is set to 1, we select the bulk region between $AB$ and the minimal cut for $AB$; this is (the spatial slice of) the {\bf entanglement wedge} $W(AB)$. If the bit is set to 0, we select the complementary bulk region, denoted $\overline{W(AB)}$. Repeating this exercise over all terms on the LHS, we recognize that the bulk can divided into $2^l$ pieces labeled by $x \in \{0,1\}^l$, which are intersections of the $W(X_i)$ or their complements; see Figure~\ref{fig:wx}. We call these bulk regions $W(x)$. Once again referring to inequality~(\ref{mmi1}), an example is:
\begin{equation}
W(101) = {W(AB)} \cap \overline{W(BC)} \cap {W(CA)}
\end{equation}
Notice that if a LHS term appears in the inequality with some multiplicity $\alpha_i \neq 1$, it only makes sense to assign it a single bit $x_i$. Putative bit strings, which assign both 0 and 1 to different copies of $S_{X_i}$ would involve the intersection $W(X_i) \cap \overline{W(X_i)} = \emptyset$. This is why $l$ counts \emph{distinct} LHS terms.

In Figure~\ref{fig:wx}, every RT surface is divided into segments by intersections with other RT surfaces. On the two sides of each segment are regions $W(x)$ and $W(x')$ for some bit strings $x, x' \in \{0,1\}^l$. Let us count how many times the segment that separates $W(x)$ from $W(x')$ contributes on the LHS of the inequality. If the $i^{\rm th}$ bit of $x$ and $x'$ is different, it means that that segment is on the RT surface that divides $W(X_i)$ from $\overline{W(X_i)}$, so it is part of the area that computes $S_{X_i}$. If the $i^{\rm th}$ bit of $x$ and $x'$ is the same, the segment is not part of $S_{X_i}$. Adding up contributions of all pairs $x, x' \in \{0,1\}^l$ to all LHS terms, we find
\begin{equation}
{\rm LHS} 
= \frac{1}{2} \sum_{x, x' \in \{0,1\}^l} |x-x'| \times \textrm{Area\,\big(RT segment between $W(x)$ and $W(x')$\big)}
\label{lhseval}
\end{equation} 
where
\begin{equation}
|x - x'| := \sum_{i=1}^l \alpha_i\, | x_i - x'_i| 
\label{xxdist}
\end{equation}
and $x_i$ is the $i^{\rm th}$ bit in string $x$. 

Proofs by contraction assemble {\bf candidate entanglement wedges} for regions $Y_j$, which appear on the RHS of the inequality. Specifically, a contraction is a map $f: x \to f(x)$ where $x \in \{0,1\}^l$ and $f(x) \in \{0,1\}^r$, with $r$ counting the total number of RHS terms. The $j^{\rm th}$ bit of $f(x)$ tells us whether or not $W(x)$ is part of the candidate entanglement wedge for $Y_j$:
\begin{equation}
W(Y_j)_{\rm candidate} = \bigcup_{x \in \{0,1\}^l \, |\, f(x)_j = 1} W(x)
\label{wcandidate}
\end{equation} 
With this ansatz for $W(Y_j)$, the {\bf candidate RT surface} for $Y_j$ consists of those segments of LHS RT surfaces, which separate $W(x)$ from $W(x')$ with $|f(x)_j - f(x')_j| = 1$. By the same argument that led to (\ref{lhseval}), the RHS computed by these candidate RT surfaces equals:
\begin{equation}
{\rm RHS}_{\rm candidate} = \frac{1}{2} \sum_{x, x' \in \{0,1\}^l} |f(x)-f(x')| \times \textrm{Area\,\big(RT segment between $W(x)$ and $W(x')$\big)}
\label{rhseval}
\end{equation} 
where
\begin{equation}
|f(x) - f(x')| := \sum_{j=1}^r | f(x)_j - f(x')_j| .
\label{ffdist}
\end{equation}
Notice that when RHS terms come with a multiplicity, $f(x)$ has separate bits for each copy. This is to allow equation~(\ref{wcandidate}) to potentially construct different candidate RT surfaces for otherwise identical RHS terms. Of course, exploiting this possibility would produce candidate RT surfaces, which are definitely unphysical. The degree to which candidate RT surfaces can be taken seriously is one of the main topics of this paper. 

Comparing (\ref{lhseval}) and (\ref{rhseval}), we see that
\begin{equation}
\forall~x, x' \in \{0,1\}^l \qquad |x - x'| \geq |f(x) - f(x')|
\label{defcontraction}
\end{equation}
guarantees that
\begin{equation}
{\rm LHS} \geq {\rm RHS}_{\rm candidate} \geq {\rm RHS},
\end{equation}
thereby proving the inequality. The relation ${\rm RHS}_{\rm candidate} \geq {\rm RHS}$ holds because the physical RHS is obtained by minimizing over all candidate right hand sides. Inequality~(\ref{defcontraction}) is the {\bf contraction condition}. Maps $f$ with this property are {\bf contractions}.

\paragraph{Boundary conditions} To prove an inequality, a contraction must satisfy certain boundary conditions. A contraction only `knows' about the inequality through the boundary conditions. The physical meaning of the boundary conditions is that they enforce the homology requirement for the candidate entanglement wedges in (\ref{wcandidate}).

The homology requirement states that the entanglement wedge of $W(Y_j)$ must contain the near-boundary region that is immediately adjacent to $Y_j$ and must exclude the near-boundary regions that are immediately adjacent to $\overline{Y_j}$. For a fundamental subsystem $A$, the $W(x)$ which is immediately adjacent to $A$ is $W(x^A)$ defined by:
\begin{equation}
(x^A)_i = \begin{cases} \,1 & {\rm if}~A \subset X_i \\ \, 0 & {\rm otherwise} \end{cases}
\label{defxai}
\end{equation}
Equation~(\ref{defxai}) means that $W(x^A)$ is adjacent to $A$ because it is contained in every $W(X_i)$ with $X_i \supset A$ and it is excluded from every $W(X_i)$ with $X_i \not\supset A$. By the same token, the candidate entanglement wedges in (\ref{wcandidate}) will satisfy the homology condition if $W(x^A)$ is part of every $W(Y_j)_{\rm candidate}$ with $A \subset Y_j$ and is excluded from all candidate wedges of $Y_j \not\supset A$. These inclusions are expressed by:
\begin{equation}
(f^A)_j = \begin{cases} \,1 & {\rm if}~A \subset Y_j \\ \, 0 & {\rm otherwise} \end{cases}
\end{equation}
In the end, the homology condition for the fundamental region $A$ takes the form:
\begin{equation}
f(x^A)  = f^A
\end{equation}
We impose such boundary conditions for every fundamental subsystem. 

\paragraph{Simplification} To verify that a map $f$ is a contraction, it is sufficient to consider pairs $(x,x')$ that differ only in a single bit. We will refer to this circumstance as a {\bf bit flip in $x$}. Considering bit flips is sufficient because we can always get from any $x$ to any $x'$ by a sequence of bit flips. If (\ref{defcontraction}) holds at every bit flip then it must also hold for $x$ and $x'$. 

We shall make frequent use of this simplification. 

\subsection{Entanglement wedge nesting}
\label{sec:ewnreview}

Minimal cuts---in the holographic context with time reversal symmetry, RT surfaces---have a simple property: If $X \subset Y$ or $X \subset \overline{Y}$ then their minimal cuts do not intersect. We refer to this statement as the {\bf entanglement wedge nesting (EWN) theorem}. Its importance for proofs by contraction is the main topic of this paper.

\begin{figure}[t]
    \centering
    \scalebox{1}[1]{\includegraphics[width=0.8\linewidth]{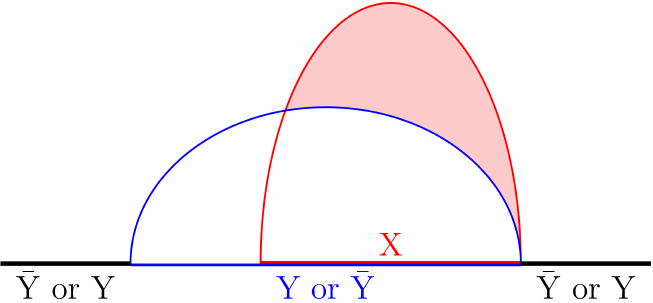}}
    \caption{Setup forbidden by the entanglement wedge nesting theorem. If both blue and red curves were minimal then combining them would make an even more minimal cut for $X$. The region highlighted in pink is the putative $W(X) \setminus W(Y)$, which would become recoverable from $\rho_Y$ only after tracing out $Y \setminus X$.}
    \label{fig:ewn}
\end{figure}

As a geometrical statement, the EWN theorem is manifestly true. Minimal cuts $\mathcal{C}_X$ and $\mathcal{C}_Y$ for a pair of nested regions $X \subset Y$ cannot intersect because if they did, we could replace the part of $\mathcal{C}_X$ that lies beyond the intersection with its counterpart from $\mathcal{C}_Y$ and obtain an even more minimal cut for $X$; see Figure~\ref{fig:ewn}. In holography, the EWN theorem holds an additional significance. The bulk region between the boundary system $Y$ and its RT surface---the spatial slice of the entanglement wedge $W(Y)$---is that portion of the bulk, which can be recovered from $\rho_Y$ alone \cite{Czech:2012bh, dongharlowwall}. This statement is known as subregion duality. If the EWN theorem did not hold, we would be in the paradoxical situation that for $Y \supset X$, tracing out $Y \setminus X$ to obtain
\begin{equation}
\rho_X = {\rm Tr}_{Y \setminus X}\, (\rho_Y)
\end{equation}
would allow one to decode an extra bulk region $W(X) \setminus W(Y)$. In other words, `forgetting' some boundary information would reveal extra information about the bulk. In bulk language, the EWN theorem has been proven using an independent argument that assumes the null energy condition \cite{Wall:2012uf}, which shows that entanglement wedge nesting is intimately related to the consistency and dynamics of bulk physics.

\subsection{Questions and organization}
\label{sec:questions}
One might expect that entanglement wedge nesting should simplify the search for contraction maps. After all, the purpose of a contraction is to produce candidate RT surfaces for the right hand side. If entanglement wedge nesting constrains RT surfaces, shouldn't it also constrain contraction maps?  We highlight this possibility as the first focal question of the paper:
\medskip

{\bf Question 1: Does wedge nesting fix entries in a contraction table?}
\medskip 

\noindent
Before answering this question, let us illustrate how this \emph{might} work. We take the monogamy of mutual information as an example.

\paragraph{Example: Monogamy of mutual information}
To prove
\begin{equation}
S_{AB} + S_{BC} + S_{CA} \geq S_A + S_B + S_C + S_{ABC}
\label{eq:mmi}
\end{equation}
by contraction, one must fill in the missing entries in Table~\ref{tab:mmi1}. The complete rows in Table~\ref{tab:mmi1} are the boundary conditions, which enforce the homology requirement in the Ryu-Takayanagi proposal. 
\begin{table}
\centering
\begin{tabular}{l ccc p{0.1cm} cccc}
  \hline
  & \multicolumn{3}{c}{$x$} & & \multicolumn{4}{c}{$f({x})$} \\
  \cline{2-4} \cline{6-9}
      & ~$AB$~ & ~$BC$~ & ~$CA$~ & & ~$A$~ & ~$B$~ & ~$C$~ & ~$ABC$~\\
  $O$~~ & 0 & 0 & 0 & & 0 & 0 & 0 & 0 \\
      & 0 & 0 & 1 & & & & & \\
      & 0 & 1 & 0 & & & & & \\
  $C$ & 0 & 1 & 1 & & 0 & 0 & 1 & 1 \\
      & 1 & 0 & 0 & & & & & \\
  $A$  & 1 & 0 & 1 & & 1 & 0 & 0 & 1 \\
  $B$  & 1 & 1 & 0 & & 0 & 1 & 0 & 1 \\
      & 1 & 1 & 1 & & & & & \\
  \hline
\end{tabular}
\caption{An incomplete contraction for proving the monogamy of mutual information. Only the boundary conditions are filled in.}
\label{tab:mmi1}
\end{table}

For entry $A$ in row $001$, the following reasoning suggests itself. The `0' in the $AB$ column means that the $001$ region is outside the entanglement wedge of $AB$. Wedge nesting says that the $AB$ entanglement wedge contains the $A$ entanglement wedge. Putting these facts together, we have:
\begin{equation}
W(001) \subset \overline{W(AB)} \subset \overline{W(A)}
\end{equation}
In the end, we have $W(001) \subset \overline{W(A)}$. If the candidate RT surfaces constructed by the contraction have anything to do with the physically realized RT surfaces, we should put a 0 in the $A$ column of row $001$.

A similar reasoning suggests how to set the $ABC$ entry in row $001$. The `1' in the $CA$ column means that the $001$ region is within the entanglement wedge of $CA$. Wedge nesting says that the $CA$ entanglement wedge is inside the $ABC$ entanglement wedge. These facts together state:
\begin{equation}
W(001) \subset W(CA) \subset W(ABC)
\end{equation}
We find $W(001) \subset W(ABC)$ so we should put a `1' in the $ABC$ column of the $001$ row.

More generally, suppose an inequality contains a term $S_V$ on the `greater than' side (LHS) and a term $S_W$ on the `less than' side (RHS). If regions $X$ and $Y$ (or their complements) are in a nested arrangement then entanglement wedge nesting suggests the following rules for filling a contraction table:
\begin{align}
\mathbf{Nesting~rules\!:}~~~~
& \color{blue}{\rm if}~X_{LHS} \supset Y_{RHS} && \color{blue}{\rm and}~~x_{X} = 0~~~~{\rm then}~~f(x)_Y=0 \nonumber \\
& \color{Orange}{\rm if}~X_{LHS} \subset Y_{RHS}&& \color{Orange} {\rm and}~~x_{X} = 1~~~~{\rm then}~~f(x)_Y = 1 \label{nestingrules} \\
& \color{red} {\rm if}~X_{LHS} \cap Y_{RHS} = \emptyset && \color{red} {\rm and}~~x_{X} = 1~~~~{\rm then}~~f(x)_Y = 0~~~~~~
\nonumber
\end{align}

Whenever possible, we will use this color scheme in highlighting instances of specific nesting rules. However, note that each rule in (\ref{nestingrules}) is equivalent to the other two under the operation of taking complements. Indeed, when we trade $X_{LHS} \to \overline{X_{LHS}}$ in the upper line, or $Y_{RHS} \to \overline{Y_{RHS}}$ in the middle line, we obtain the bottom line. 

When we apply these rules to fill out Table~\ref{tab:mmi1}, we obtain Table~\ref{tab:mmi2}, which is the correct and complete contraction table. Evidently, rules~(\ref{nestingrules}) work in proving the monogamy of mutual information. What is more, they write out the entire proof for us. Judging by this example, the answer to Question~1 seems to be affirmative. 

\begin{table}
\centering
\begin{tabular}{l ccc p{0.1cm} cccc}
  \hline
  & \multicolumn{3}{c}{$\vec{x}$} & & \multicolumn{4}{c}{$\vec{y} = f(\vec{x})$} \\
  \cline{2-4} \cline{6-9}
      & ~$AB$~ & ~$BC$~ & ~$CA$~ & & ~$A$~ & ~$B$~ & ~$C$~ & ~$ABC$~\\
  $O$~~ & 0 & 0 & 0 & & 0 & 0 & 0 & 0 \\
      & {\color{blue} \framebox{0}} & 0 & {\color{Orange} \framebox{1}} & & {\color{blue} \framebox{0}} & {\color{blue} 0} & {\color{blue} 0} & {\color{Orange} \framebox{1}} \\
      & 0 & 1 & 0 & & {\color{blue} 0} & {\color{blue} 0} & {\color{blue} 0} & {\color{Orange} 1} \\
  $C$ & 0 & 1 & 1 & & 0 & 0 & 1 & 1 \\
      & 1 & 0 & 0 & & {\color{blue} 0} & {\color{blue} 0} & {\color{blue} 0} & {\color{Orange} 1} \\
  $A$  & 1 & 0 & 1 & & 1 & 0 & 0 & 1 \\
  $B$  & 1 & 1 & 0 & & 0 & 1 & 0 & 1 \\
      & {\color{red} \framebox{1}} & 1 & 1 & & {\color{red} 0} & {\color{red} 0} & {\color{red} \framebox{0}} & {\color{Orange} 1} \\
  \hline
\end{tabular}
\caption{The complete contraction for proving the monogamy of mutual information. We box three specific applications of line~1 ({\color{blue} blue}), line~2 ({\color{Orange} orange}) and line~3 ({\color{red} red}) in nesting rules~(\ref{nestingrules}). All other entries in the table can be obtained by the same reasoning (also color-coded).}
\label{tab:mmi2}
\end{table}

As it turns out, this example is spectacularly misleading. 
\medskip

{\bf Answer to Question 1: No! }
\medskip

\noindent
The reason for the exclamation mark is that, in fact, {\bf proofs by contraction \emph{necessarily} break nesting rules}. Fixing entries in a contraction table based on entanglement wedge nesting is not merely unhelpful; it makes finding the contraction impossible. Based on the current knowledge of the holographic entropy cone \cite{HernandezCuenca:2019wgh, Czech:2022fzb, Hernandez-Cuenca:2023iqh, Czech:2023xed}, the monogamy of mutual information is the \emph{unique} exception in this regard.

In this paper, entries in a contraction table, which do not conform to rules~(\ref{nestingrules}), are referred to as {\bf nesting violations}. The main claim in Section~\ref{sec:violation} is that every proof by contraction of every (known) holographic entropy inequality contains at least one nesting violation, with Table~\ref{tab:mmi2} being the only exception. In fact, in senses that we quantify in Sections~\ref{sec:maxdense} and \ref{sec:375}, proofs of known maximally tight holographic entropy inequalities contain {\it as many nesting violations as possible} and occur nearly {\it as commonly as possible}. By the criterion used in Section~\ref{sec:375}, 1854 out of the 1868 inequalities considered therein conform to this rule of thumb. 

We emphasize that nesting violations do not invalidate proofs by contraction. That is because a contraction is supposed to produce any one candidate RT surface whereas nesting is only guaranteed for the actual, physical surface. Nevertheless, we find it surprising that proofs by contraction not only tolerate nesting violations but, in fact, rely on them for survival. 

These findings pose many questions about contraction maps as a general strategy of proof. For starters, if no proofs by contraction respect rules~(\ref{nestingrules}) then:
\medskip

{\bf Question 2: What role does nesting play in proofs by contraction?}
\medskip 

To outline and probe this role, we discuss our recent proof of the two infinite families of holographic entropy inequalities, which are associated with tessellations of the torus and the projective plane \cite{Czech:2023xed}. As we review below, the tessellations are an encoding of entanglement wedge nesting. The utility of the tessellations in constructing and proving the inequalities tells us that nesting plays a role in the holographic entropy cone, but what is it? We shall analyze the mechanics and motivation behind the inequalities' proof by contraction to gain insight. 

Qualitatively speaking, we find that {the freedom and difficulty in finding contraction maps rests in designing a scheme to break wedge nesting}. The argument involves an interesting interplay with the topology of the underlying kinematic space \cite{Czech:2015qta}---the space of subregions, which are being probed by the inequality. This analysis comprises Section~\ref{sec:study} of our paper. In addition, Appendix~\ref{app:12} applies the same analysis to another inequality, which is outside the families studied in Reference~\cite{Czech:2023xed}. 

Section~\ref{sec:discussion} is a discussion, which is concerned with broader implications.

\subsection{Notation}

We will often find it useful to divide all regions participating in an inequality into two classes $A_i$ and $B_j$, with $1 \leq i \leq m$ and $1 \leq j \leq n$. When we do so, the division will include the purifying region, such that the state on $(A_1 \ldots A_m B_1 \ldots B_n)$ is pure. The indices $i$ and $j$ are assumed to be cyclically ordered (mod $m$) and (mod $n$), i.e. $i = 1 \equiv m+1$. 

With reference to this bidivision, it is further useful to define consecutively ordered $k$-tuples of $A$- and $B$-regions:
\begin{equation}
A_i^{(k)} = A_i A_{i+1} \ldots A_{i+k-1} \qquad {\rm and} \qquad B_j^{(k)} = B_j B_{j+1} \ldots B_{j+k-1}
\end{equation}
Finally, when $m$ and $n$ (numbers of $A$-type and $B$-type regions) are odd, we will use the shorthand:
\begin{equation}
A_i^\pm = A_i^{((m\pm 1)/2)} \qquad {\rm and} \qquad B_j^\pm = B_j^{((n\pm 1)/2)}
\end{equation}
These objects represent the smallest consecutive majorities and the largest consecutive minorities among $A$-regions and $B$-regions.

\section{Every proof by contraction contains a nesting violation}
\label{sec:violation}

At the time of writing, known holographic entropy inequalities can be divided into three classes: 
\begin{itemize}
\item The toric family \cite{Czech:2023xed}. It is parameterized by integers $m$ and $n$, which are both odd.
\begin{equation}
\sum_{i=1}^m \sum_{j=1}^n S_{A_i^+ B_j^-}
\geq 
\sum_{i=1}^m \sum_{j=1}^n S_{A_i^- B_j^-} \,+ S_{A_1A_2\ldots A_m}
\label{toricineqs}
\end{equation}
The recent paper \cite{Bao:2024vmy} proposed a further generalization of (\ref{toricineqs}) but it is not known how many of the new inequalities are independent. 
\item The projective plane family \cite{Czech:2023xed}. It is parameterized by one integer $m = n$, which can be even or odd.
\begin{equation} 
\!\!\!
\frac{1}{2} 
\sum_{j=1}^{m-1} \sum_{i=1}^{m}
\left(S_{A_i^{(j)} B_{i+j-1}^{(m-j)}} + S_{A_i^{(j)} B_{i+j}^{(m-j)}} \right)
\,+\, (m-1)\,S_{A_1 A_2 \ldots A_m}
\geq
\sum_{i,j=1}^{m} S_{A_i^{(j-1)} B_{i+j-1}^{(m-j)}}
\label{rp2ineqs}
\end{equation}  
\item 1868 other inequalities \cite{HernandezCuenca:2019wgh,Hernandez-Cuenca:2023iqh}, for which no organizing pattern is known.
\end{itemize}
We discuss each class of inequalities in turn.  

\subsection{The simplest example}
\label{sec:example}
To appreciate how and why nesting violations are necessary in proofs by contraction (beyond monogamy), it is clarifying to inspect one example in detail. We take an example from the toric family. It turns out that the $(m,1)$ subfamily of the toric inequalities, which has previously been called `cyclic' or `dihedral' \cite{Bao:2015bfa,Czech:2023xed}, obscures certain aspects of the problem. 
Therefore, we consider the $(m,n) = (3,3)$ toric inequality.

This inequality, which was first found in \cite{Czech:2022fzb}, reads:
\begin{align} 
   S_{A_1 A_2 B_1} \,+\, S_{A_2 A_3 B_1}  \,+\, S_{A_3 A_1 B_1} 
& \phantom{~~\geq~~ +.}\!
     S_{A_3 B_1} \,+\, S_{A_1 B_1} \,+\, S_{A_2 B_1} 
\nonumber \\
    +\phantom{.} S_{A_1 A_2 B_2} \,+\, S_{A_2 A_3 B_2} \,+\, S_{A_3 A_1 B_2} 
&  ~~\geq~~  +\phantom{.}\! 
     S_{A_3 B_2} \,+\, S_{A_1 B_2} \,+\, S_{A_2 B_2}  ~~~+\, S_ {A_1 A_2 A_3}  
\nonumber \\    
     +\phantom{.} S_{A_1 A_2 B_3} \,+\, S_{A_2 A_3 B_3}  \,+\, S_{A_3 A_1 B_3}  
& \phantom{~~\geq~~} +\phantom{.}\! 
          S_{A_3 B_3} \,+\, S_{A_1 B_3} \,+\, S_{A_2 B_3}    \label{i5}
\end{align}
The terms of inequality~(\ref{i5}) are displayed as two-dimensional arrays on each side, plus a special term on the right hand side. These arrays, whose rows and columns can be periodically permuted, are the reason why the inequality is associated with the torus. 

A contraction map for proving this inequality takes an arrangement of 0's and 1's assigned to left hand side terms, and maps it to an arrangement of 0's and 1's assigned to right hand side terms. Using the same graphical arrangement as in (\ref{i5}), one example of the mapping is:
\begin{equation}
\begin{tabular}{|ccc|}
\hline
1 & 1 & 0 \\
1 & 1 & 0 \\
1 & 1 & 0 \\
\hline
\end{tabular}
~~\longrightarrow~~
\begin{tabular}{|ccc|c}
\cline{1-3}
0 & 0 & 1 & \\
\cline{4-4}
0 & 0 & 1 & \multicolumn{1}{c|}{1} \\
\cline{4-4}
0 & 0 & 1 & \\
\cline{1-3}
\end{tabular}
\label{a2boundary}
\end{equation}
This specific assignment is shared by all contraction maps that prove (\ref{i5}) because it is a boundary condition. Equation~(\ref{a2boundary}) is the boundary condition for region $A_2$: we have 1's on all terms that contain $A_2$ and 0's on all terms that do not.

One virtue of the arrangement of terms in (\ref{i5}) is that it makes applications of nesting rules~(\ref{nestingrules}) transparent and mechanical:
\begin{itemize}
\item According to the first (blue) line in (\ref{nestingrules}), a 0 on the left hand side mandates 0's in the other two columns of the same row on the right hand side.
\item According to the third (red) line in (\ref{nestingrules}), a 1 on the left hand side mandates 0's in the other two rows of the same column on the right hand side.
\end{itemize} 
We color-code two examples of these rules below:
\begin{equation}
\begin{tabular}{|ccc|}
\hline
\color{red} 1 & 1 & \color{blue} 0 \\
1 & 1 & 0 \\
1 & 1 & 0 \\
\hline
\end{tabular}
~~\longrightarrow~~
\begin{tabular}{|ccc|c}
\cline{1-3}
\color{blue} 0 & \color{blue} 0 & 1 & \\
\cline{4-4}
\color{red} 0 & 0 & 1 & \multicolumn{1}{c|}{1} \\
\cline{4-4}
\color{red} 0 & 0 & 1 & \\
\cline{1-3}
\end{tabular}
\label{a2boundary-color}
\end{equation}
One take-away message from~(\ref{a2boundary-color}) is that if nesting rules (\ref{nestingrules}) were to be applied uniformly, they would significantly constrain the contraction map. We claim that the nesting rules routinely over-constrain contractions.

In the case of the $(3,3)$ toric inequality, the over-constraining is manifest. Consider what the contraction can do one bit away from the boundary condition~(\ref{a2boundary}): 
\begin{equation*}
\begin{tabular}{|ccc|}
\hline
1 & 1 & 0 \\
1 & 1 & 0 \\
1 & 1 & \color{red} 1 \\
\hline
\end{tabular}
~~\longrightarrow~~
\begin{tabular}{|ccc|c}
\cline{1-3}
? & ? & ? & \\
\cline{4-4}
? & ? & ? & \multicolumn{1}{c|}{?} \\
\cline{4-4}
? & ? & ? & \\
\cline{1-3}
\end{tabular}
\end{equation*}
If the nesting rules were applied, they would determine most of the array on the right hand side:
\begin{equation}
\begin{tabular}{|ccc|}
\hline
1 & 1 & 0 \\
1 & 1 & 0 \\
1 & 1 & \color{red} 1 \\
\hline
\end{tabular}
~~\xrightarrow{~~\textrm{nesting rules}~~}~~
\begin{tabular}{|ccc|c}
\cline{1-3}
0 & 0 & \color{red} 0 & \\
\cline{4-4}
0 & 0 & \color{red} 0 & \multicolumn{1}{c|}{?} \\
\cline{4-4}
0 & 0 & ? & \\
\cline{1-3}
\end{tabular}
\label{offa2boundary}
\end{equation}
However, the array on the right hand side of~(\ref{offa2boundary}) is at least distance 2 away from the right hand side of (\ref{a2boundary}). Therefore, assignment~(\ref{offa2boundary}) does not satisfy the requirements of a contraction.
 
\paragraph{Observations} We offer a few qualitative observations about this example:
\begin{itemize}
\item The unavoidable nesting violation occurs near a boundary condition. 
\item In this example---and in proofs of all toric inequalities---nesting rules~(\ref{nestingrules}) tend to put many 0's on the right hand side of a contraction table. (This is the effect of the first and the third line in (\ref{nestingrules}).) On the other hand, boundary conditions require a minimum number of 1's on the right hand side. This produces a tension between entanglement wedge nesting and the boundary conditions. 
\end{itemize}
We expect that contraction proofs of maximally tight holographic entropy inequalities somehow balance these two rationales---nesting versus boundary conditions. Our analysis shows that that balance is not achievable with zero nesting violations (except in the case of monogamy). We elaborate on this in Section~\ref{sec:unknown}.

\subsection{The toric family} 
\label{sec:toric}
The argument from the previous subsection applies to the whole toric family \cite{Czech:2023xed}. We copy the toric inequalities below for the reader's convenience:
\begin{equation}
\sum_{i=1}^m \sum_{j=1}^n S_{A_i^+ B_j^-}
\geq 
\sum_{i=1}^m \sum_{j=1}^n S_{A_i^- B_j^-} \,+ S_{A_1A_2\ldots A_m}
\qquad\qquad \textrm{($m$, $n$~odd)}
\label{toricineqs2}
\end{equation}

\paragraph{Graph}
To streamline the analysis---and for convenience in later sections---we organize the terms in the sums graphically. 
The graphical scheme is designed so that nesting rules~(\ref{nestingrules}) have a similar effect as in the $(m,n) = (3,3)$ example. We associate left hand side and right hand side terms of the inequality with features of a graph, which is embedded on a torus.

The graph is drawn according to the following rules:
\begin{itemize}
\item Terms in the sum on the right hand side are vertices (nodes) of the graph.
\item The graph is embedded on a torus. Terms on the left hand side are faces of the embedded graph.
\item A face $S_X$ is incident to a node $S_Y$ if and only if $X$ and $Y$ are in a nesting relationship. (This is sometimes expressed by saying that $X$ and $Y$ do not cross.)
\end{itemize}
The last point means that, by construction, the effect of nesting rules will be local on the graph. It will involve a face and its incident vertices, or a vertex and its incident faces. 

Let us assume $m,n > 1$. (We return to the case $n=1$ briefly at the end of the section.) Applying the above rules to inequalities~(\ref{toricineqs2}) produces a graph embedded on a torus, which has the following features:
\begin{itemize}
\item All faces are squares. That is, every left hand side term is in a nesting relationship with four right hand side terms, which are:
\begin{equation}
A_i^+ B_j^- ~ \sim ~ A_i^- B_j^-,~~ A_{i+1}^- B_j^-,~~ A_{i+(m+1)/2}^- B_{j+(n-1)/2}^-,~~ A_{i+(m+1)/2}^- B_{j+(n+1)/2}^- 
\end{equation}
\item All vertices are four-valent. That is, every term in the sum on the right hand side is in a nesting relationship with four left hand side terms:
\begin{equation}
A_i^- B_j^- ~ \sim ~
A_i^+ B_j^-,~~ A_{i-1}^+ B_j^-,~~ A_{i + (m-1)/2}^+ B_{j+(n-1)/2}^-,~~ A_{i + (m-1)/2}^+ B_{j+(n+1)/2}^-
\end{equation}
\item The special term $S_{A_1 A_2 \ldots A_m}$ does not appear in the graph because it is not in a nesting relationship with any other term. This is a consequence of assuming $m,n > 1$ for otherwise we have $A_i^+ \subset A_1 A_2 \ldots A_m$.
\item Other than the special term $S_{A_1 A_2 \ldots A_m}$, each other term in the inequality appears exactly once in the graph as either a vertex or a face.
\end{itemize}
The first two points above imply that a toric inequality is represented by a square tessellation of the torus.

\begin{figure}
		\centering
		$\begin{array}{lp{1mm}r}
		\includegraphics[width=0.347\linewidth]{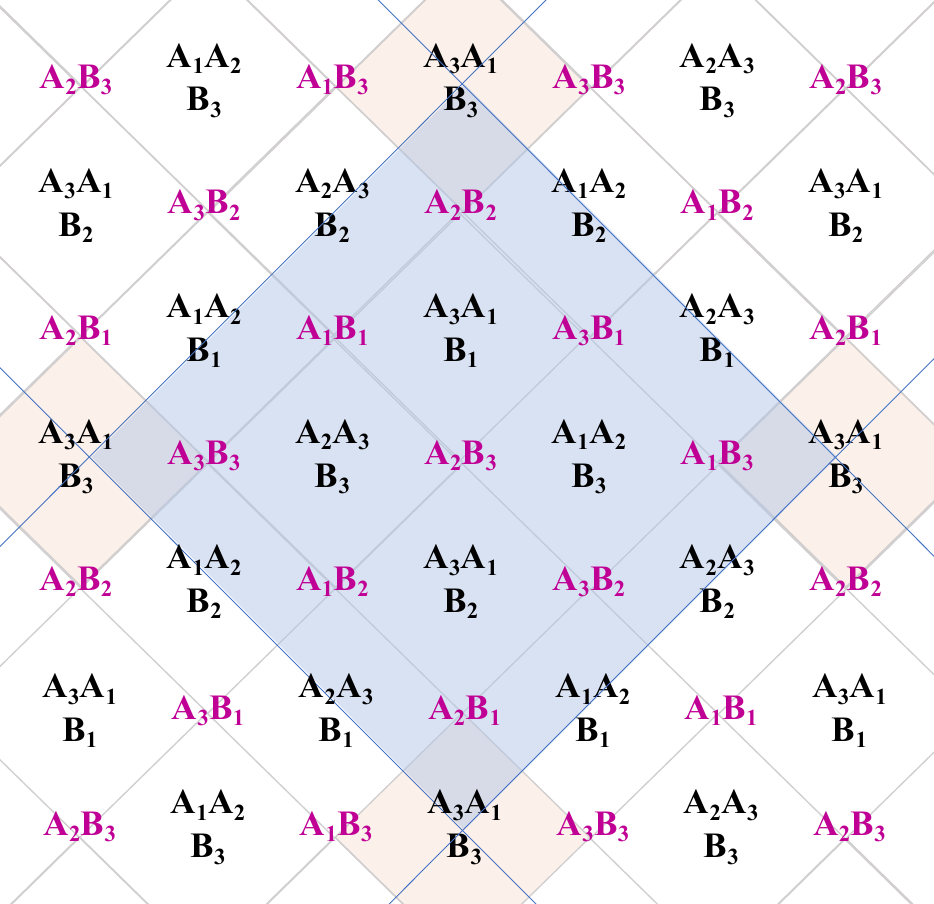} & &
		\includegraphics[width=0.621\linewidth]{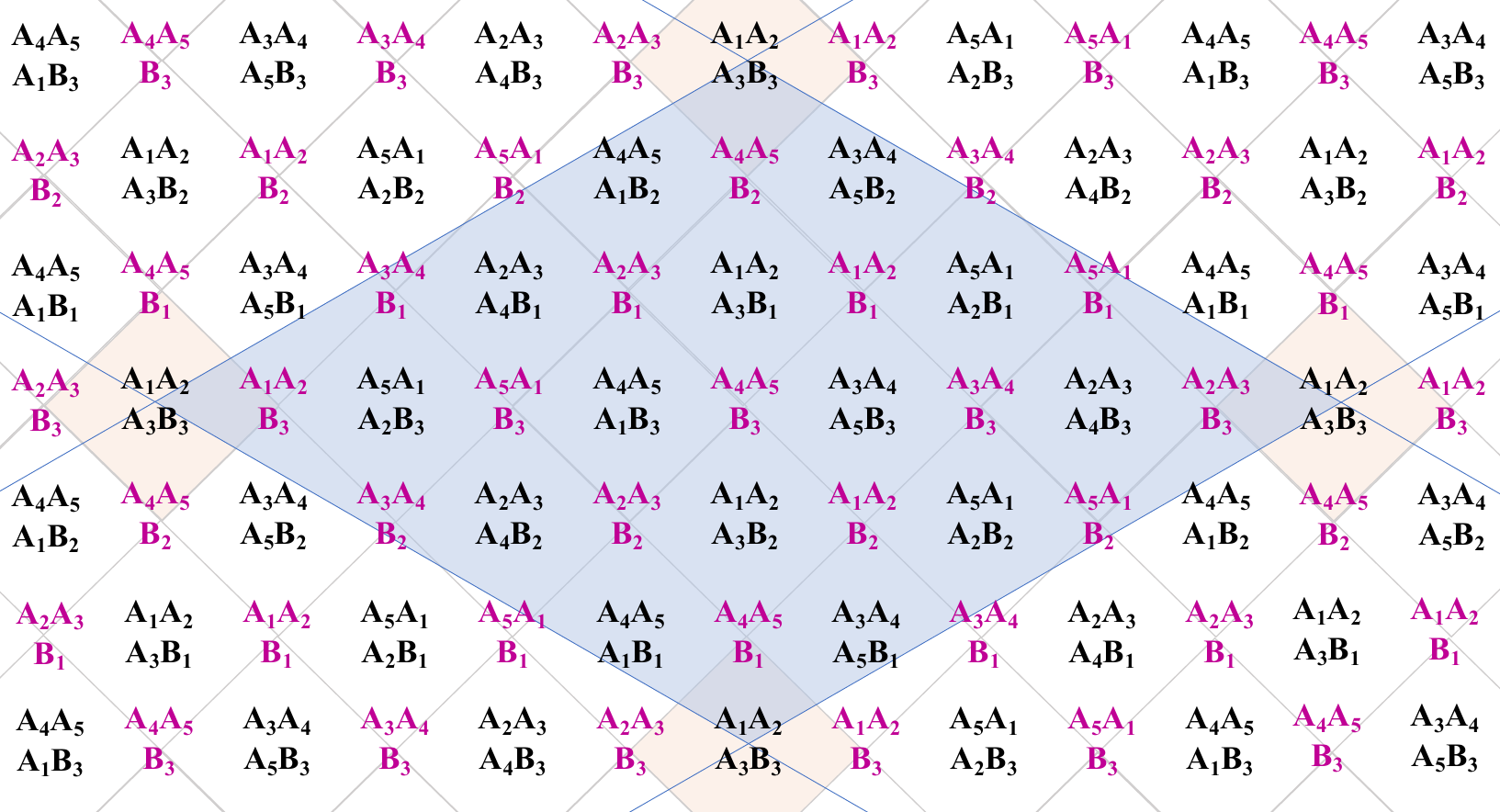}
		\end{array}$
	\caption{Graphical representations of the $(m,n) = (3,3)$ (left panel) and $(m,n) = (5,3)$ (right panel) toric inequalities. The graphs descend from plane tessellations under periodic identifications. In both panels we highlight one fundamental domain and a foursome of identical faces.} 
\label{fig:lattices}
\end{figure}

To keep the discussion transparent, let us fix conventions. First, we write the toric inequalities precisely as in (\ref{toricineqs2}); that is we never trade any term $S_V$ for its complement $S_{\overline{V}}$. Second, we orient the neighborhood of every vertex and face in the graph in the way shown below:
\begin{equation}
\begin{array}{c}
\includegraphics[width=0.6\textwidth]{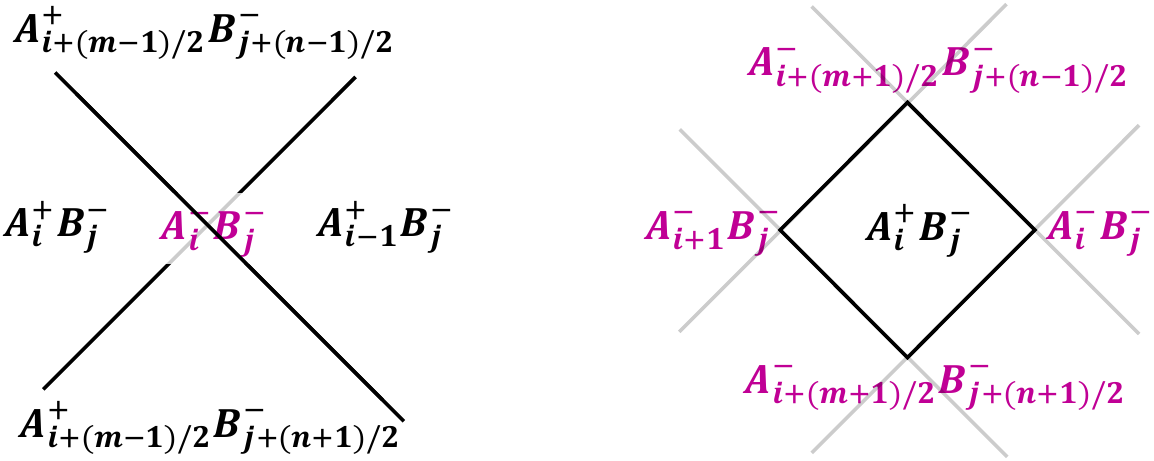}
\end{array}
\label{fig:toruslocal}
\end{equation}
These assumptions unambiguously set all choices of convention for displaying toric inequalities as graphs. The resulting graphs for $(m,n) = (3,3)$ and $(5,3)$ are shown in Figure~\ref{fig:lattices}. We encourage the reader to compare the tableaux from Section~\ref{sec:example} with the $(3,3)$ graph in Figure~\ref{fig:lattices}.

\paragraph{Edges in the graph are conditional mutual informations} The graph defined above comes with a bonus. It makes clear that inequalities~(\ref{toricineqs2}) describe a particular improvement on strong subadditivity. To see this, use the obvious fact that an edge has two endpoints and is adjacent to two faces. In other words, every edge is canonically identified with two LHS terms and two RHS terms of the inequality. 

An edge, which goes from top left to bottom right is associated with:
\begin{equation}
I_{\rm edge} 
= S_{A_i^+ B_j^-}\, + S_{A_{i+1}^- B_j^+}\, - S_{A_{i+1}^- B_j^-} - S_{A_{i}^+ B_j^+} 
= I(A_i : B_{j+(n-1)/2}\, |\, A_{i+1}^- B_j^-)
\end{equation}
For an edge, which goes from bottom left to top right, the associated terms are:
\begin{equation}
I_{\rm edge} 
= S_{A_i^+ B_{j}^-}\, + S_{A_{i}^- B_{j}^+}\, - S_{A_{i}^- B_{j}^-} - S_{A_{i}^+ B_{j}^+}
= I(A_{i+(m-1)/2} : B_{j+(n-1)/2}\, |\, A_i^- B_j^-)
\end{equation}
Each vertex and each face are incident to two right-going and two left-going edges. Therefore, summing up all left-going (alternatively, all right-going) edges double counts all vertices and faces. After doing the sum and inserting the special term $S_{A_1 A_2 \ldots A_m}$ that is not represented graphically, we find that (\ref{rp2ineqs2}) can be rewritten as:
\begin{equation}
\frac{1}{2} \sum_{i = 1}^m \sum_{j=1}^n I(A_i : B_{j+(n-1)/2}\, |\, A_{i+1}^- B_j^-) \geq S_{A_1 A_2 \ldots A_m}
\label{cmirewrite}
\end{equation}

Each $I(A_i : B_{j+(n-1)/2}\, |\, A_{i+1}^- B_j^-)$ is nonnegative by strong subadditivity. Rewriting (\ref{cmirewrite}) presents the toric inequalities as an improvement over a toroidal collection of strong subadditivities. We emphasize that the improvement term $S_{A_1 A_2 \ldots A_m}$ is the sole reason why the toric inequalities are not trivial.

\subsubsection{Nesting rules in toric inequalities} 
\begin{figure}[H]
\centering
\includegraphics[width=0.5\textwidth]{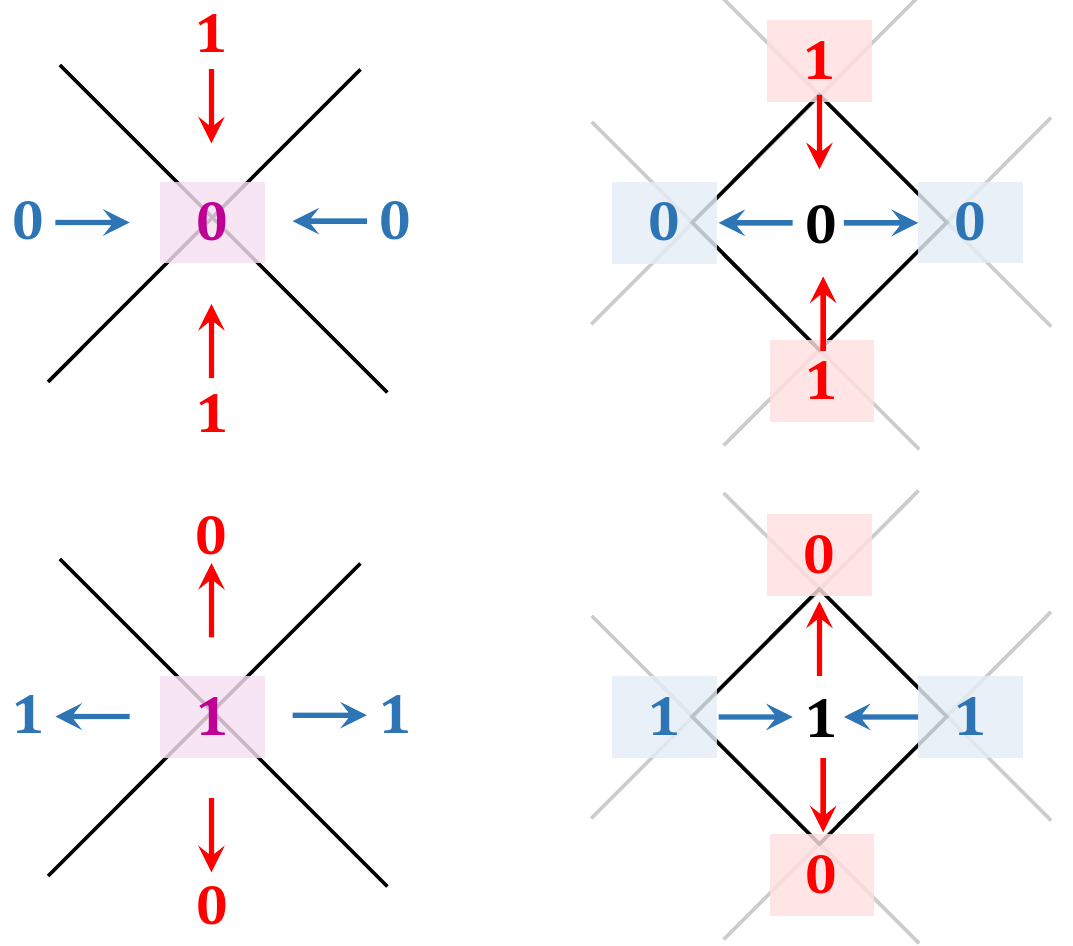}
\caption{The action of nesting rules on graph representations of toric inequalities, color coded as in (\ref{nestingrules}). Arrows represent logical implications. The arrows, which originate from `1' on a vertex are related to rules~(\ref{nestingrules}) by contrapositive. }
\label{fig:implications}
\end{figure}

A contraction map takes an assignment of 0/1 to LHS terms (faces), and produces an assignment of 0/1 to RHS terms (vertices plus the additional term $S_{A_1 A_2 \ldots A_m}$). We shall draw the 0's and 1's directly over vertices and faces of the graph. With the above conventions in place, nesting rules~(\ref{nestingrules}) act as in Figure~\ref{fig:implications}. 

It so happens that the middle (orange) line in (\ref{nestingrules}) makes no appearance on the graphs of toric inequalities with $m,n > 1$. It is amusing to think of the arrows in Figure~\ref{fig:implications}---that is, logical implications---as field lines. Then a 1 on a vertex acts like a source whereas a 0 on a vertex acts like a sink.
\medskip

\paragraph{Lemma} If any arrangement displayed on the left below appears in a contraction map then the contraction contains a nesting violation.
\begin{align}
\includegraphics[width=0.2\textwidth]{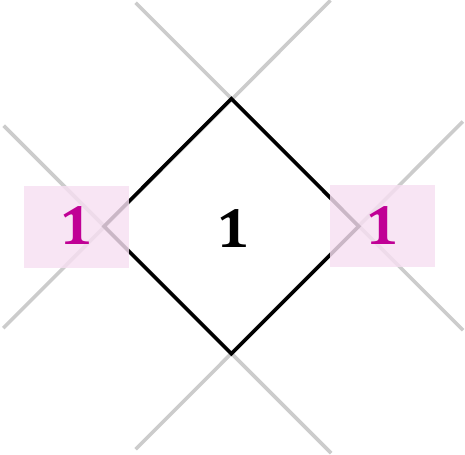} 
& \raisebox{0.095\textwidth}{$~~\xleftrightarrow{~~\textrm{dist. on RHS at least 2}~~}~~$}
\includegraphics[width=0.2\textwidth]{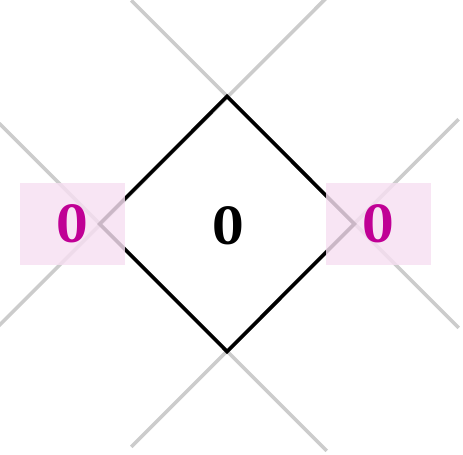} 
\raisebox{0.095\textwidth}{$~~\xleftarrow{~~\textrm{nesting rules}~~}~~$}
\label{whyviolate1}
\\
\includegraphics[width=0.2\textwidth]{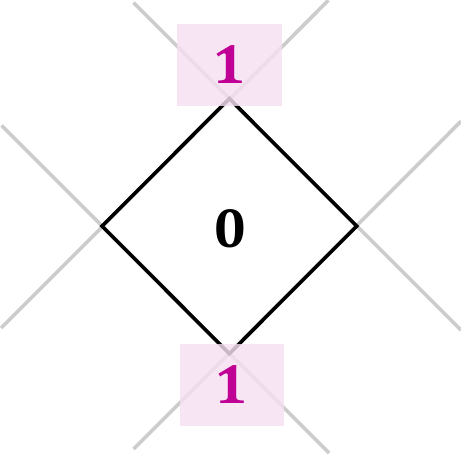} 
& \raisebox{0.095\textwidth}{$~~\xleftrightarrow{~~\textrm{dist. on RHS at least 2}~~}~~$}
\includegraphics[width=0.2\textwidth]{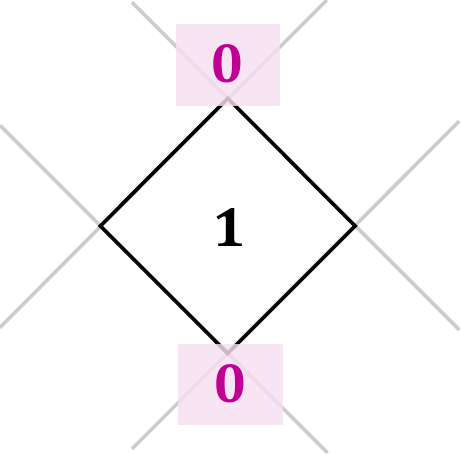}
\raisebox{0.095\textwidth}{$~~\xleftarrow{~~\textrm{nesting rules}~~}~~$}
\label{whyviolate2}
\end{align}
The reason is manifest in equations~(\ref{whyviolate1}-\ref{whyviolate2}): the arrangements on the left are `protected' by nesting rules. This means that, if nesting rules were respected, changing the 0/1 on the face would alter the assignment on at least two neighboring vertices and preclude the map from being a contraction. We saw an example of this in equation~(\ref{offa2boundary}). 

The lemma will allow us to efficiently detect nesting violations in a broad class of inequalities, not just in toric inequalities. 
\medskip 

\begin{figure}[t]
		\centering
		$\begin{array}{lr}
		\includegraphics[width=0.426\linewidth]{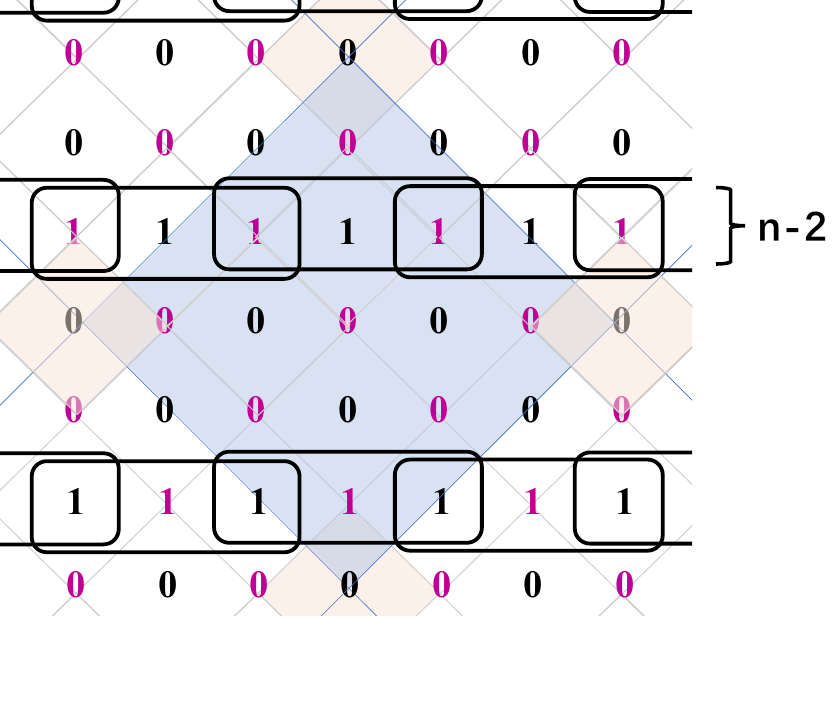} &
		\includegraphics[width=0.55\linewidth]{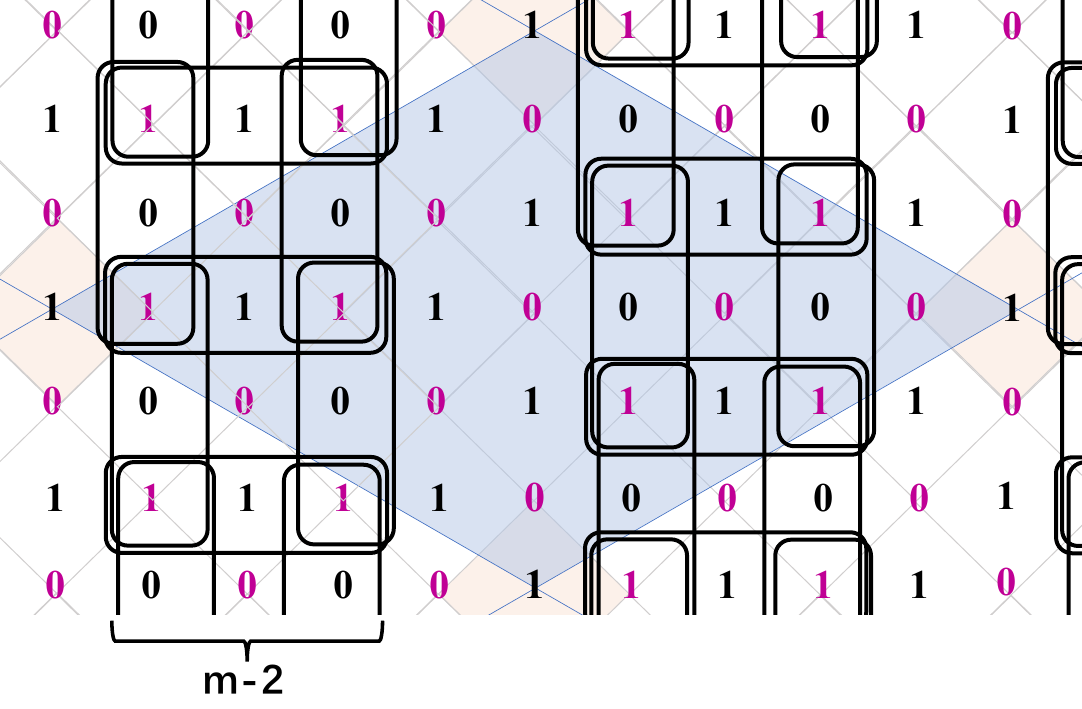}
		\end{array}$
	\caption{Boundary conditions for toric inequalities, with highlighted instances of nesting violation-inducing configurations from equations~(\ref{whyviolate1}-\ref{whyviolate2}). Displayed are boundary conditions for $B_j$ in the $(3,3)$ inequality (left) and for $A_i$ in the $(5,3)$ inequality (right).} 
\label{fig:bctoric}
\end{figure}

\paragraph{All boundary conditions satisfy the premise of the lemma} We show examples of boundary conditions of toric inequalities in Figure~\ref{fig:bctoric}. The configurations in (\ref{whyviolate1}-\ref{whyviolate2}), which occasion nesting violations, arrange themselves into $(m-1)/2$ (or, respectively, $(n-1)/2$) lines, which wrap around the torus and connect to one another. They are always there for $m,n \geq 3$. Therefore, all contractions, which prove the $m,n \geq 3$ toric inequalities, contain nesting violations.

\begin{figure}[t]
		\centering
		\includegraphics[width=0.80\linewidth]{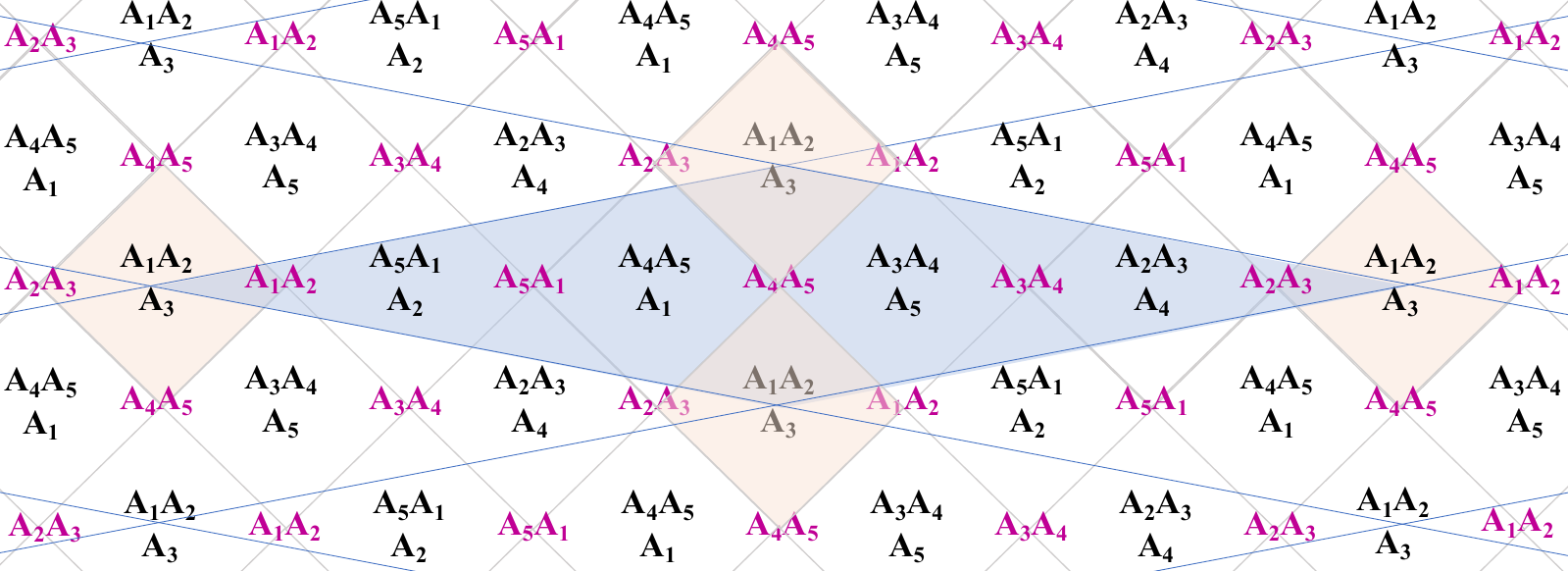} \\
		\vspace*{4mm}
	\includegraphics[width=0.80\linewidth]{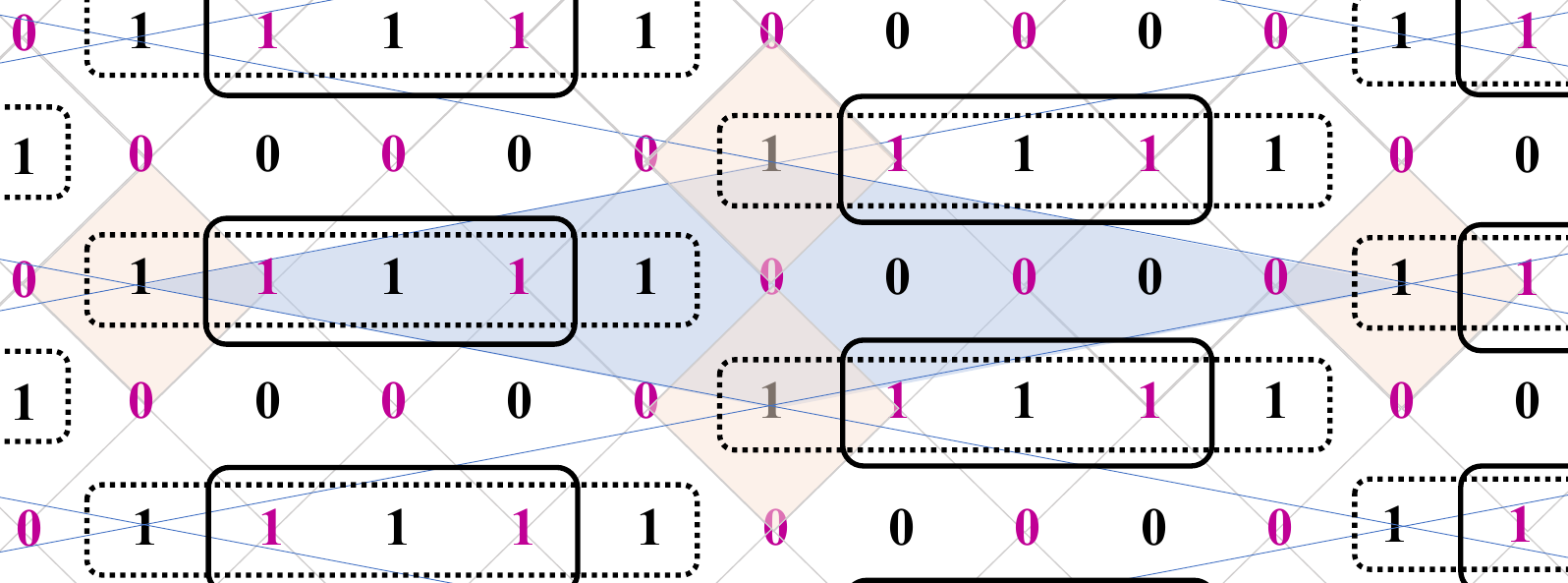}
	\caption{Graphical representation and a boundary condition for the dihedral inequalities ($n=1$), here shown for $m=5$. This graph does not reflect the nesting rule in the middle (orange) line in (\ref{nestingrules}). We highlight arrangements~(\ref{whyviolate1}), which induce nesting violations, as well as the larger sequences 1-1-1-1-1 on sites face-vertex-face-vertex-face, which are referenced in the main text.} 
\label{fig:dihedral}
\end{figure}

\paragraph{Dihedral inequalities} The preceding analysis is also valid for the dihedral inequalities ($n=1$). If we apply the same graphical rules as we did to the other toric inequalities, we obtain Figure~\ref{fig:dihedral}. While the figure does not make manifest applications of the middle (orange) line in rules~(\ref{nestingrules}), it turns out that proofs by contraction respect that specific nesting rule. Therefore, we focus on the other two nesting rules, that is the top and bottom lines in (\ref{nestingrules}). Those are accurately reflected by the graph in Figure~\ref{fig:dihedral}. 

The arrangement in equation~(\ref{whyviolate2}) does not appear in Figure~\ref{fig:dihedral}. There are places where the graph shows a vertical vertex-face-vertex sequence with 1-0-1, but in such instances the two 1's represent the same point on the torus. Thus, scenario (\ref{whyviolate2}) does not occur in proofs of dihedral inequalities. Arrangement~(\ref{whyviolate1}), on the other hand, does occur. 

It is interesting to contrast Figure~\ref{fig:dihedral} with the spectacularly misleading example from Section~\ref{sec:questions}. What makes the contraction for the monogamy of mutual information (the dihedral inequality with $m=3$) uniquely nesting violation-free? In Figure~\ref{fig:dihedral}, which displays the case $m=5$, the requisite vertex-face-vertex pattern 1-1-1 (equation~\ref{whyviolate1}) appears as part of a longer sequence 1-1-1-1-1, which begins and ends on faces. That is, we have a horizontal sequence of $m=5$ 1's that live on sites face-vertex-face-vertex-face. If $m=3$, the analogue of Figure~\ref{fig:dihedral} contains instead three 1's on sites face-vertex-face, which does not contain a subsequence vertex-face-vertex. This is the reason why in dihedral inequalities scenario~(\ref{whyviolate1}) requires $m \geq 5$.

\subsection{The projective plane family}
\label{sec:rp}
Here we apply the same technique to the projective plane inequalities, copied for the reader's convenience below:
\begin{equation} 
\!\!\!
\frac{1}{2} 
\sum_{j=1}^{m-1} \sum_{i=1}^{m}
\left(S_{A_i^{(j)} B_{i+j-1}^{(m-j)}} + S_{A_i^{(j)} B_{i+j}^{(m-j)}} \right)
+ (m-1)\, S_{A_1 A_2 \ldots A_m}
\geq
\sum_{i,j=1}^{m} S_{A_i^{(j-1)} B_{i+j-1}^{(m-j)}}
\label{rp2ineqs2}
\end{equation}  
We represent the nesting relationships among terms of (\ref{rp2ineqs2}) using a graph, this time embedded on the projective plane. We then verify that boundary conditions for the contraction always contain patterns (\ref{whyviolate1}-\ref{whyviolate2}), which are sufficient conditions for nesting violations.

\paragraph{Graph} Before drawing the graph, we highlight one feature of~(\ref{rp2ineqs2}): terms on the left hand side form complementary pairs. For example, the term $S_{A_i^{(j)} B_{i+j}^{(m-j)}}$ equals $S_{A_{i+j}^{(m-j)} B_{i}^{(j)}}$, which also appears in the sum. This is because the two regions are complements of one another and our discussion assumes the state on $A_1 A_2 \ldots A_m B_1 B_2 \ldots B_m$ to be pure. 

We shall soon see that the symmetry between terms $S_X$ and $S_{\overline{X}}$ becomes an inversion symmetry in the graph representation of (\ref{rp2ineqs2}). The projective plane will arise as a quotient under this inversion symmetry. For now, we choose to write the inequalities in the form~(\ref{rp2ineqs2})---that is, using on the left hand side \mbox{$\tfrac{1}{2}(S_X + S_{\overline{X}})$} rather than simply $S_{\overline{X}}$ or $S_X$---in order to manifest the symmetry under inversion (taking complements).

Proceeding as in the case of the toric inequalities, we represent nesting relations among terms as incidence relations between vertices and faces. In analogy to (\ref{fig:toruslocal}), we find that a generic term in (\ref{rp2ineqs2}) has four neighbors as shown below:
\begin{equation}
	\!\!\!\!\!\begin{array}{lp{0.1mm}r}
		\includegraphics[width=0.44\linewidth]{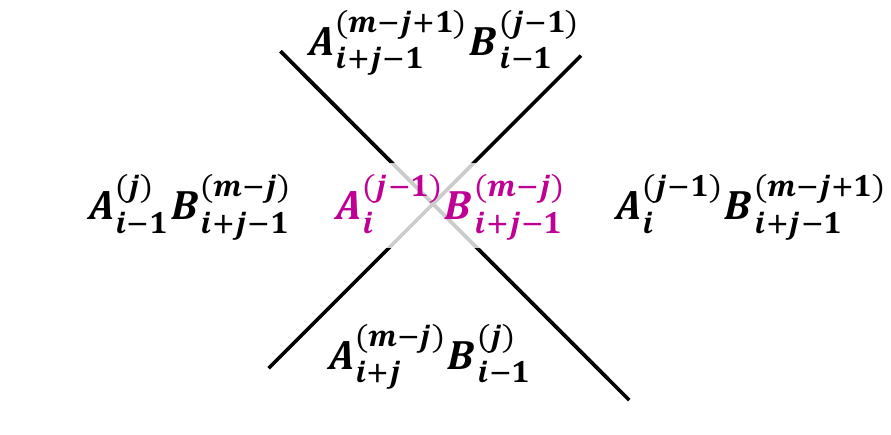} & &
		\includegraphics[width=0.44\linewidth]{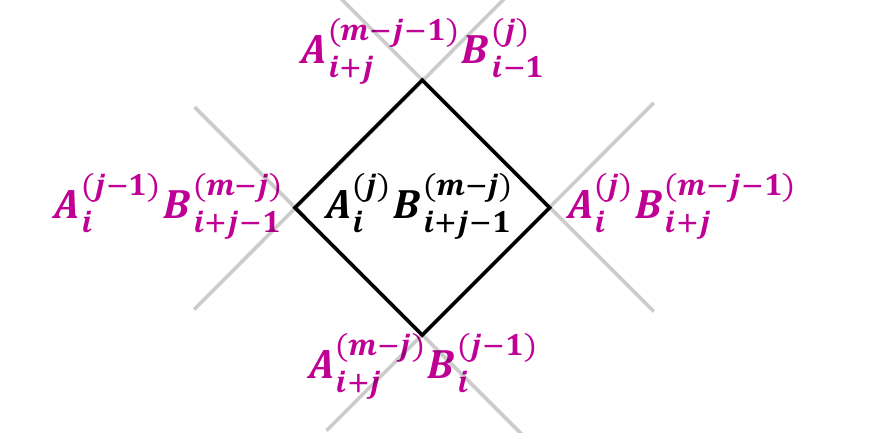} \\
		\includegraphics[width=0.44\linewidth]{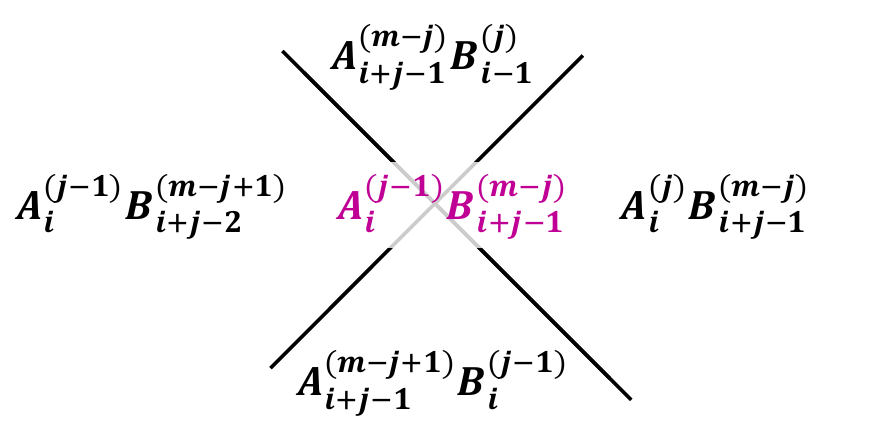} & &
		\includegraphics[width=0.44\linewidth]{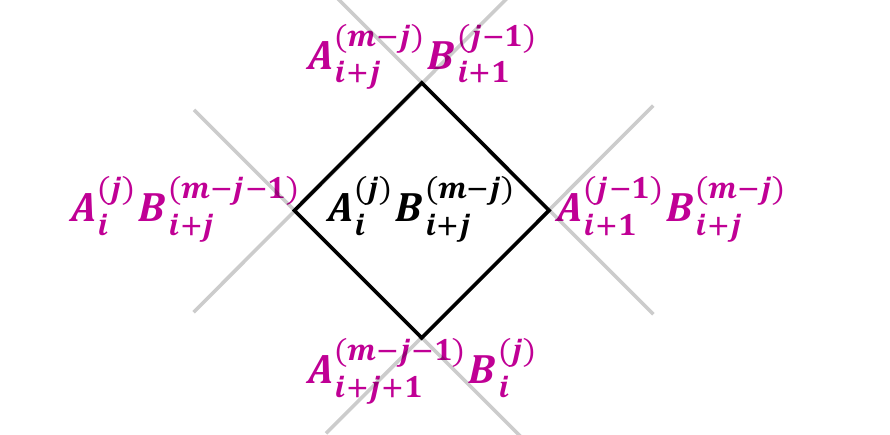} 
		\end{array}
\label{rp2local}
\end{equation}
In the left panels, the neighborhood of the same vertex is drawn in two different ways. They are related to one another by a reflection across the diagonal. As we discuss momentarily, this symmetry is important in realizing the graph as a tiling of the projective plane.

Non-generic circumstances concern the special term $S_{A_1 A_2 \ldots A_m} = S_{B_1 B_2 \ldots B_m}$. It is in a nesting relationship with $2m$ distinct right hand side terms $S_{A_i^{(m-1)}}$ and $S_{B_i^{(m-1)}}$. Conversely, those $2m$ terms are in nesting relationships with three left hand side terms, one of which is $S_{A_1 A_2 \ldots A_m} = S_{B_1 B_2 \ldots B_m}$. Even for those $2m$ vertices, however, the graphical representation in (\ref{rp2local}) remains valid. When $j=1$ or $j=m$, two of the four incident faces shown in (\ref{rp2local}) become $S_{A_1 A_2 \ldots A_m} = S_{B_1 B_2 \ldots B_m}$.

\begin{figure}[t]
		\centering
		$\begin{array}{lr}
		\includegraphics[width=0.462\linewidth]{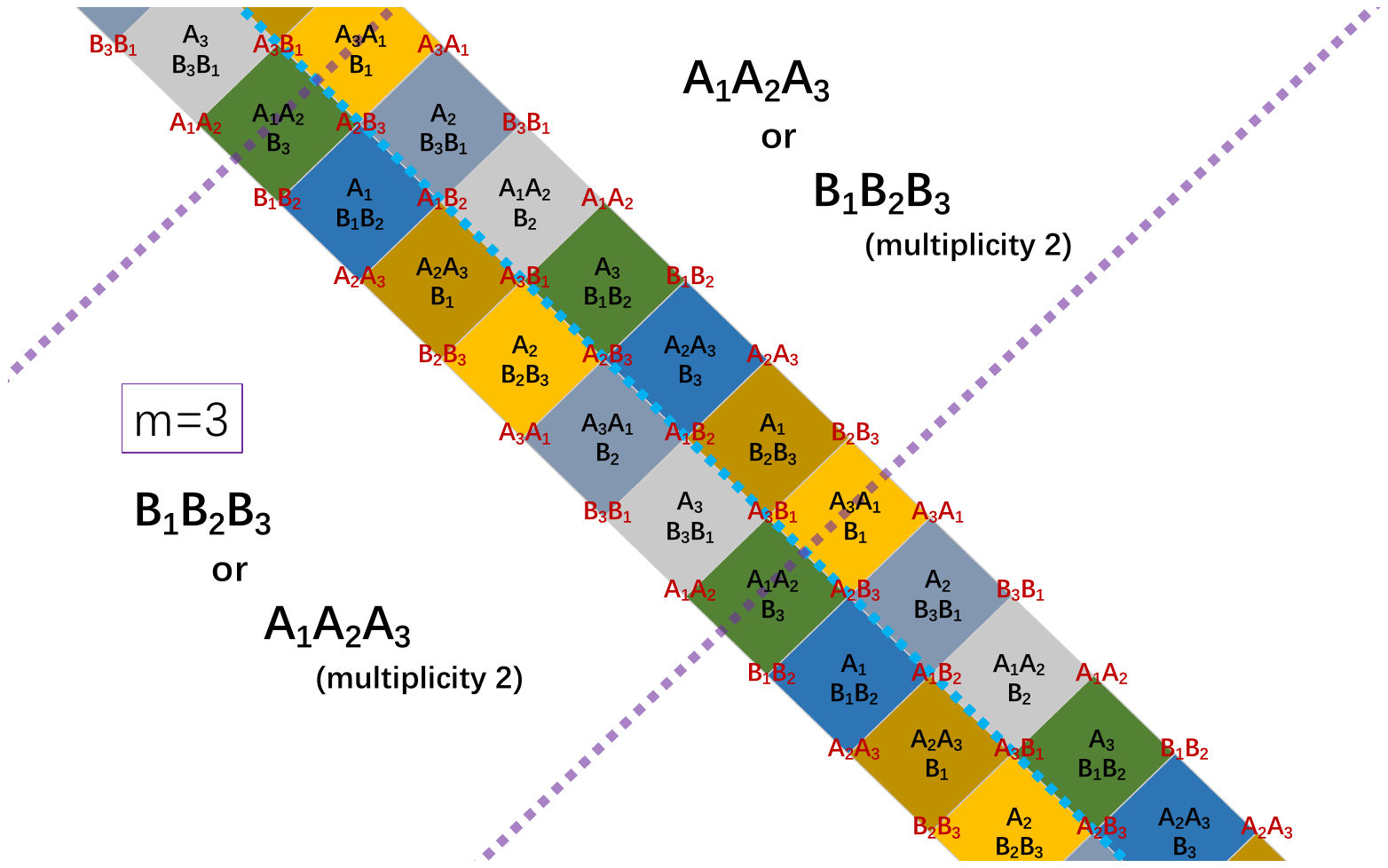} &
		\includegraphics[width=0.525\linewidth]{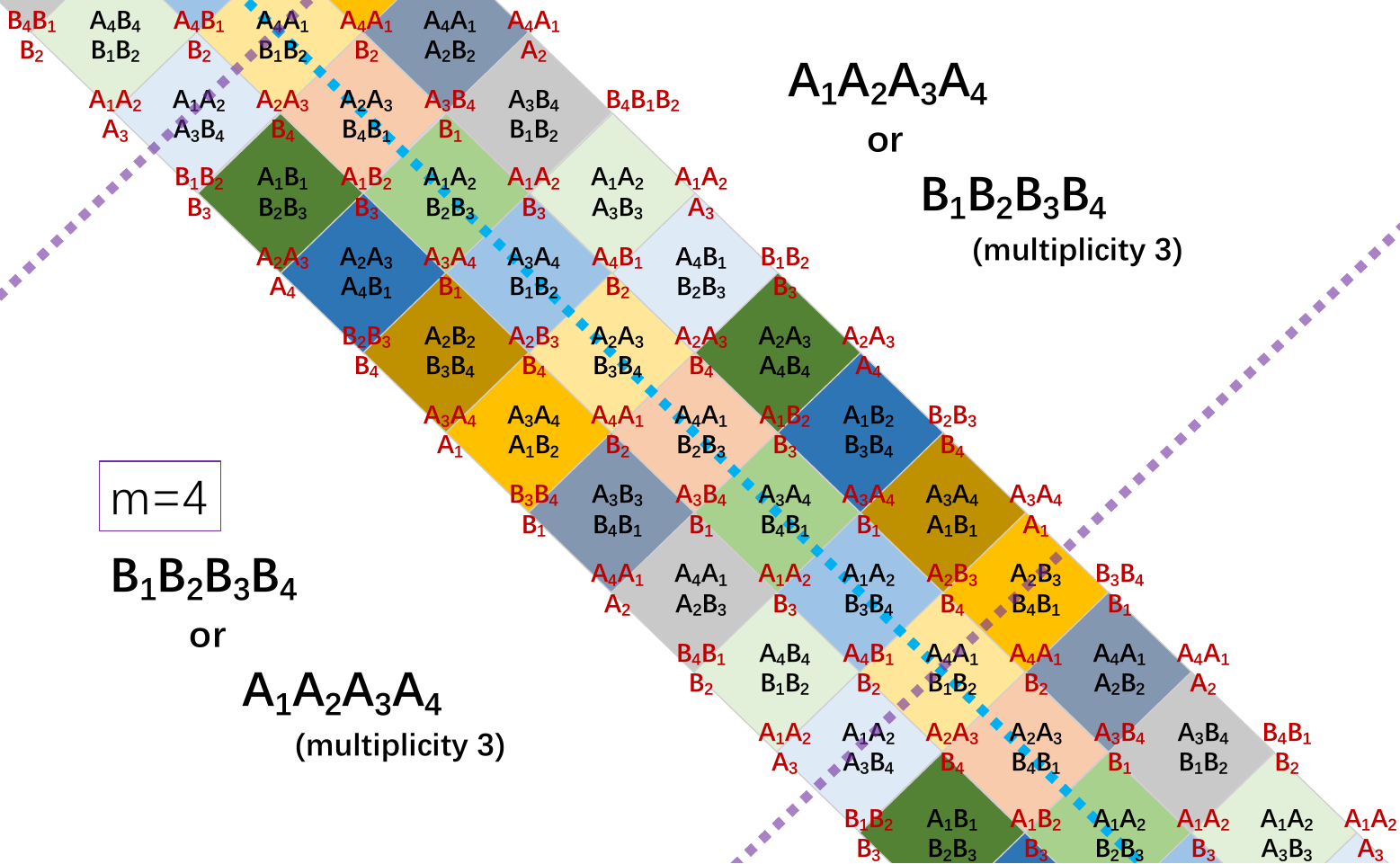}
		\end{array}$
	\caption{Graphs that represent inequalities~(\ref{rp2ineqs2}) for $m=3$ and $m=4$. The region between the purple dotted lines is topologically a sphere. The blue dotted line (the central axis of the strip), which is the equator of the sphere, is actually a cross-cap. A single fundamental domain, which is enclosed by the purple and blue lines, is the projective plane.} 
\label{fig:rp}
\end{figure}

We now put together the graphical elements in (\ref{rp2local}) like a puzzle. 
If we keep adding to the puzzle forever, we will arrive at an infinite strip shown in Figure~\ref{fig:rp}. On the other hand, if we close the graph after each puzzle piece in (\ref{rp2local}), for all $i$ and $j$, has been used exactly once, we arrive at a cylinder. The special terms $S_{A_1 A_2 \ldots A_m} = S_{B_1 B_2 \ldots B_m}$ cap the cylinder from the top and bottom, making it into a sphere. The sphere is visible in Figure~\ref{fig:rp} as the region between purple lines.

Notice that the graph on the sphere covers two copies of inequality~(\ref{rp2ineqs2}). On the left hand side (counting faces), the graph contains one full sum over $S_{A_i^{(j)} B_{i+j-1}^{(m-j)}}$ and $S_{A_i^{(j)} B_{i+j}^{(m-j)}}$, in which every independent term $S_X$ is accompanied by $S_{\overline{X}}$. On the right hand side (vertices), every independent term $S_{A_i^{(j-1)} B_{i+j-1}^{(m-j)}}$ appears twice---once as shown in the upper line of (\ref{rp2local}), and once as shown in the lower line. We would like a graphical description of the inequality, in which every independent term appears once.

To find it, we observe that identical terms on the sphere appear at locations, which are related by an inversion.\footnote{Here is how to recognize the inversion. On the sphere, the central axis of the strip in Figure~\ref{fig:rp} is periodic with period $2m$. In the same figure, the symmetry that relates identical terms is a glide reflection with offset $m$, i.e. a reflection across the central axis of the strip followed by a translation by $m$ squares. A glide reflection along a periodically identified axis whose offset equals half the period is an inversion. This means that the central axis of the strip is a cross-cap on the sphere, rendering the projective plane. For an alternative discussion, which arrives at the projective plane by gluing a M{\"o}bius strip to a disk, see \cite{Czech:2023xed}.} Therefore, a single copy of inequality~(\ref{rp2ineqs2}), with every independent term appearing once, lives on the sphere modded out by an inversion---that is, the projective plane. Notice that the central axis of the strip---the dotted blue line in Figure~\ref{fig:rp}---is mapped by the inversion to itself so it is a cross-cap.

\paragraph{Edges in the graph are conditional mutual informations} As in the toric inequalities, edges in the graph are conditional mutual informations. An edge that is drawn from top left to bottom right in our convention is associated with: 
\begin{align}
I_{\rm edge} 
& = S_{A_i^{(j)} B_{i+j-1}^{(m-j)}} + S_{A_i^{(j-1)} B_{i+j-1}^{(m-j+1)}} - S_{A_i^{(j-1)} B_{i+j-1}^{(m-j)}} - S_{A_i^{(j)} B_{i+j-1}^{(m-j+1)}} \nonumber \\
& = I( A_{i+j-1} : B_{i-1} \, |\, A_i^{(j-1)} B_{i+j-1}^{(m-j)})
\label{rp2edge}
\end{align}
To flesh out the consequences of equating edges with conditional mutual informations, it is convenient to rewrite (\ref{rp2ineqs2}) by adding $S_{A_1 A_2 \ldots A_m}$ on both sides:
\begin{equation} 
\!\!\!
\frac{1}{2} 
\sum_{i,j=1}^{m}
\left(S_{A_i^{(j)} B_{i+j-1}^{(m-j)}} + S_{A_i^{(j)} B_{i+j}^{(m-j)}} \right)
\geq
\sum_{i,j=1}^{m} S_{A_i^{(j-1)} B_{i+j-1}^{(m-j)}} + S_{A_1 A_2 \ldots A_m}
\label{rp2ineqs3}
\end{equation}  
Summing up all edges from (\ref{rp2edge}) precisely double counts the sums in (\ref{rp2ineqs3}). In the end, we find that the projective plane inequalities can be rewritten in the form:
\begin{equation}
\frac{1}{2} \sum_{i,j = 1}^m I( A_{i+j-1} : B_{i-1} \, |\, A_i^{(j-1)} B_{i+j-1}^{(m-j)}) \,\geq\, S_{A_1 A_2 \ldots A_m}
\label{rp2ineqs4}
\end{equation}
In inequalities~(\ref{rp2ineqs4}), a collection of strong subadditivities assemble a discretized projective plane and gain an improvement by $S_{A_1 A_2 \ldots A_m}$.

\paragraph{Nesting rules and boundary conditions}
We chose the layout and orientation in (\ref{rp2local}) in a purposeful way, so that:
\begin{itemize}
\item Nesting rules (\ref{nestingrules}) act on the graphs of the projective plane inequalities according to Figure~\ref{fig:implications}---the same way they do on the graphs of the toric inequalities.
\end{itemize}
This means that our lemma still applies and (\ref{whyviolate1}-\ref{whyviolate2}) continue to act as sufficient conditions for a nesting violation. 

\begin{figure}[t]
		\centering
		$\begin{array}{lr}
		\includegraphics[width=0.465\linewidth]{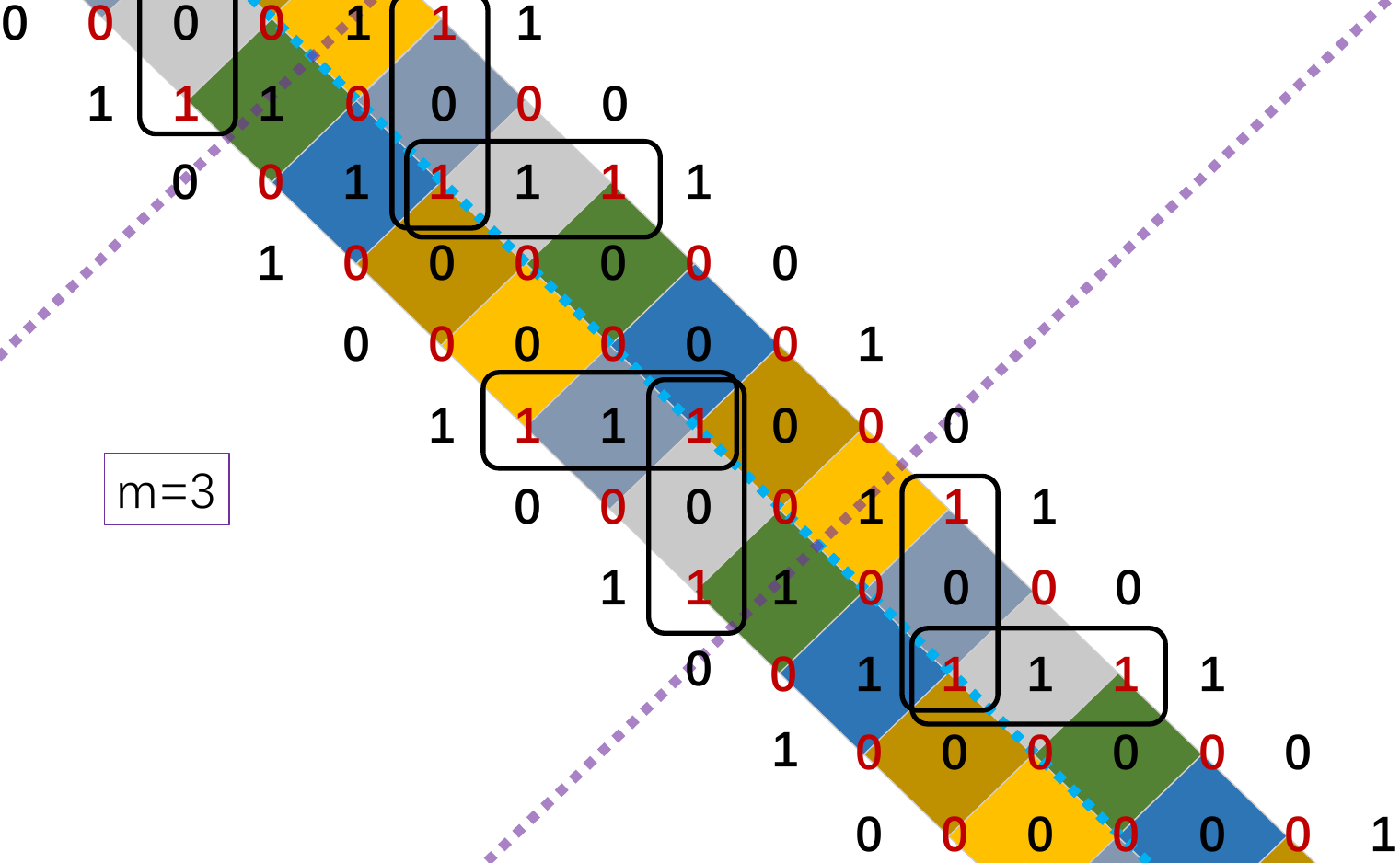} &
		\includegraphics[width=0.520\linewidth]{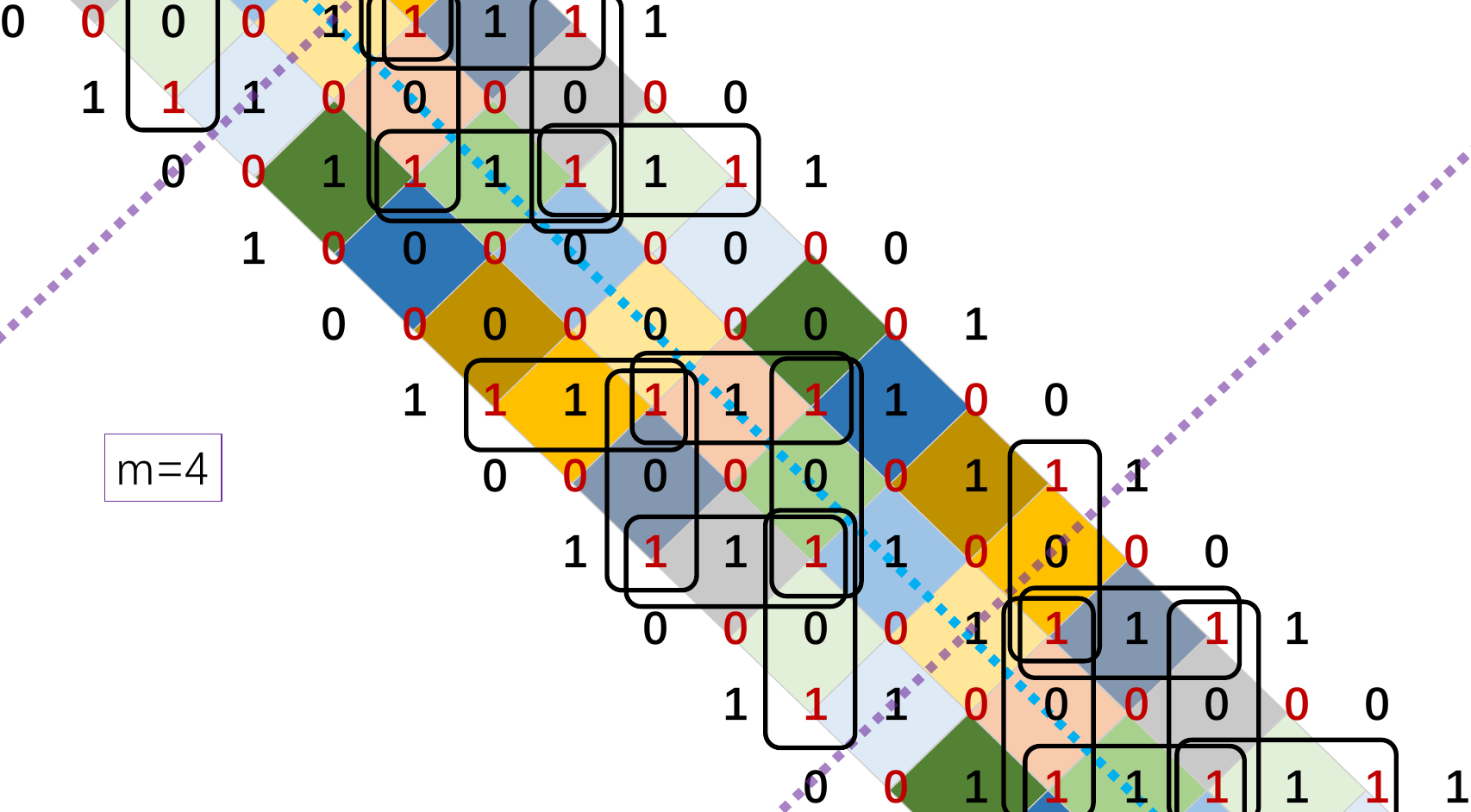}
		\end{array}$
	\caption{Boundary conditions for contractions, which prove inequalities~(\ref{rp2ineqs2}) for $m=3,4$. We highlight arrangements (\ref{whyviolate1}) and (\ref{whyviolate2}), which guarantee that the contraction map contains nesting violations.} 
\label{fig:rpbc}
\end{figure}

We display the boundary condition in Figure~\ref{fig:rpbc}. (Boundary conditions for all regions look the same; they are related by a symmetry of the inequality.) The figure contains configurations (\ref{whyviolate1}) and (\ref{whyviolate2}), which establishes that every proof by contraction of (\ref{rp2ineqs}) contains nesting violations.

\paragraph{Nesting violations in boundary conditions are maximally dense} 
\label{sec:maxdense}
%
A striking common feature of Figures~\ref{fig:bctoric} and \ref{fig:rpbc} is that configurations~(\ref{whyviolate1}-\ref{whyviolate2}) appear with the maximal possible density allowed by the graph. To make this assertion rigorous, consider the following problem:
\begin{itemize}
\item Populate an infinite chessboard with 0s and 1s, as is done in graphs in Sections~\ref{sec:toric} and \ref{sec:rp}, subject to two conditions:
\begin{itemize}
\item Implications shown in Figure~\ref{fig:implications} (nesting rules) hold.
\item Configurations~(\ref{whyviolate1}-\ref{whyviolate2}) (that is, sufficient conditions for nesting violations in proofs by contraction) occur with maximal density per unit area. 
\end{itemize}
\end{itemize} 
It is clear that the solution is to cover the infinite chessboard with alternating horizontal lines of 1s and 0s. (For now we leave this as an exercise for the reader, but we return to this point in Section~\ref{sec:howtoric}.) Figure~\ref{fig:bctoric} and Figure~\ref{fig:rpbc} exhibit this pattern over maximal patches of the graph. However, for global reasons, the pattern cannot cover the whole finite graph and a defect is unavoidable. 

Maximally packed combinations of (\ref{whyviolate1}-\ref{whyviolate2})---that is, alternating horizontal lines of 1s and 0s with a single defect line---are precisely the boundary conditions of inequalities~(\ref{toricineqs}) and (\ref{rp2ineqs}). In Figure~\ref{fig:bctoric}, the defect arises because the number of $A$- and $B$-type regions is odd. In Figure~\ref{fig:rpbc}, the defect arises because alternating lines of 1s and 0s are not consistent with the inversion symmetry. 

What is the meaning of the fact that nesting violations occur with maximal density? In colloquial terms, it is not as if the inequalities reluctantly tolerate one or two oddball nesting violations in their proofs by contraction. Rather, it appears as if they are selected by a principle that boundary conditions should \emph{maximize} nesting violations. We inspect this possibility further in the next subsection and in the Discussion.

\subsection{Other known inequalities: Nesting violations are common}
\label{sec:375}
Reference~\cite{Hernandez-Cuenca:2023iqh} recently announced 1866 new holographic entropy inequalities. In addition, there are two known holographic entropy inequalities from \cite{HernandezCuenca:2019wgh}, which are not part of the toric and $\mathbb{RP}^2$ families. In this section we verify that their proofs by contraction, without exception, also necessarily contain nesting violations.

In fact, in 1854 of the 1868 inequalities considered here, a stronger statement holds. Thus far, we have looked for unremovable nesting violations in neighborhoods of boundary conditions, using configurations (\ref{whyviolate1}) and (\ref{whyviolate2}) as smoking guns. To conclude that a nesting violation is unavoidable, we needed one occurrence of (\ref{whyviolate1}) or (\ref{whyviolate2}) in {\it at least one} boundary condition per inequality. A more extreme possibility is that not one but \emph{all} boundary conditions in a given inequality contain (\ref{whyviolate1}) or (\ref{whyviolate2}). Among the 1868 inequalities considered in this subsection, 1854 have this property. Among the fourteen that evade it, each has only one boundary condition without (\ref{whyviolate1}) or (\ref{whyviolate2}); see `Results' below.

Once again, it appears as if the inequalities are designed by some principle, which demands as many nesting violations as possible. Here `as many as possible' refers to the spread of nesting violations among boundary conditions, not as a density on a graph. We do not consider the density of violations on a graph in this subsection because most of the 1868 inequalities do not admit useful graphical representations according to the rules spelled out in Sections~\ref{sec:toric} and \ref{sec:rp}. (One example that does admit a graph representation is discussed in Appendix~\ref{app:12}.)

Graph models being unavailable, we instead design an ad hoc technique for detecting unavoidable nesting violations:

\paragraph{Technique}
We start with an observation, which makes the analysis more transparent: to each boundary condition is associated a canonical balanced\footnote{Balanced means that every fundamental region appears a net zero times in the inequality.} form of the inequality. Indeed, in the literature on holographic inequalities one typically designates a region $O$ to be the purifier and interprets the inequality as pertaining to general (pure or mixed) states on $\overline{O}$---the complement of $O$. When doing so, one writes every term in the inequality in the way that excludes $O$ and the resulting expression is always balanced \cite{Hubeny:2018ijt}. To illustrate how and why boundary conditions define balanced formulations of every inequality, we write the monogamy of mutual information in four ways---each associated to one boundary condition:
\begin{align} 
O: &~~~~~S_{AB} + S_{BC} + S_{CA} ~\geq~ S_{A\phantom{BC}} + S_{B\phantom{AC}} + S_{C\phantom{AB}} + S_{ABC} \nonumber\\
A: &~~~~~S_{OC} + S_{BC} + S_{BO} ~\geq~ S_{BCO} + S_{B\phantom{AC}} + S_{C\phantom{AB}} + S_{O} 
\nonumber \\
B: &~~~~~S_{OC} + S_{AO} + S_{CA} ~\geq~ S_{A\phantom{BC}} + S_{ACO} + S_{C\phantom{AB}} + S_{O} 
\\
C: &~~~~~S_{AB} + S_{AO} + S_{BO} ~\geq~ S_{A\phantom{BC}} + S_{B\phantom{AC}} + S_{ABO} + S_{O} \nonumber
\end{align}

The usefulness of the above observation is as follows. When the inequality is written in the way canonically associated with a boundary condition, the latter takes a simple form: all zeroes, because every term excludes the purifier. This fact simplifies our search for configurations, which implicate nesting violations. For every boundary condition, we first bring the inequality to the associated canonical form (set the underlying region to be the purifier) and then inspect the configuration `all zeroes.'

With this starting point, let us study what happens when one LHS zero is flipped to a one. After imposing the nesting rules, a 1 on the LHS carries two types of implications: from the middle (orange) line and from the bottom (red) line of~(\ref{nestingrules}). The orange nesting rule mandates certain bits to \emph{become} 1s while the red nesting rule requires certain bits to \emph{remain} at zero. The latter requirement does nothing to bound Hamming distances between the boundary condition and its neighbors. Only the middle (orange) line in (\ref{nestingrules}) bounds distances away from the `all zeroes' configuration. 

This reasoning indicates that we can bound distances between the `all-zeroes' boundary condition and other configurations simply by counting applications of the orange line in (\ref{nestingrules}). Those occur whenever a LHS term $X$ is a subset of a RHS term $Y \supset X$. There is, however, one additional subtlety, which is best illustrated by an example.

\paragraph{Example of subtlety} This is inequality number 13 in the list provided by \cite{Hernandez-Cuenca:2023iqh}:
\begin{align}
S_{AB} \,+\, S_{ABD} \,+\, S_{ACD} \,+\, S_{ADE}\,  & +\, S_{BCD} \,+\, S_{BDE} \,+\, S_{CEF} \,+\, S_{DEF} \nonumber \\
& \geq \label{ineq13} \\
S_A \,+\, S_B \,+\, S_C \,+\, S_{AD} \,+\, S_{BD} \, & +\, S_{DE} \,+\, S_{EF} \,+\, S_{ABCD} \,+\, S_{ABDE} \,+\, S_{CDEF}
\nonumber
\end{align}
As we discussed, the boundary condition for the unnamed purifier is zeroes for all terms. (In the tables below, we mark the rows that contain that boundary condition with $O$.) Proceeding as we did in Section~\ref{sec:example}, we ask where the contraction goes when we flip the $AB$ bit from 0 to 1:
\begin{equation*}
\begin{array}{l||c|c|c|c|c|c|c|c|c|c}
O & & 0  & 0 & 0 & 0 & 0 & 0 &0 & 0 & \\
    & & \color{Orange}\framebox{1} & 0 & 0 & 0 & 0 & 0 &0 & 0 & \\
{\rm LHS}~~ & & ~{AB}~ & ~{ABD}~ & ~{ACD}~ & ~{ADE}~ & ~{BCD}~ & ~{BDE}~ & {CEF} & {DEF} & \\
\hline
\hline
{\rm RHS}~~ & ~A~ & B & C & {AD} & {BD} & {DE} & {EF} & ~{ABCD}~ & ~{ABDE}~ & ~{CDEF}~ \\
O & 0 & 0 & 0 & 0 & 0 & 0 & 0 & 0 & 0 & 0 \\
    & ? & ? & ? & ? & ? & ? & ? & \color{Orange}\framebox{1} & \color{Orange}\framebox{1} & ? 
\end{array}
\end{equation*}
Because $AB \subset ABCD$ and $AB \subset ABDE$, nesting rules~(\ref{nestingrules}) demand these two entries to become 1. The Hamming distance on the right hand side is at \mbox{least 2}. We might conclude that flipping the $AB$ bit in the `all-zeroes' boundary condition leads to a nesting violation. 

In fact, it does not. This is because $AB$ is also contained inside another LHS term: $AB \subset ABD$. Therefore, when the $AB$ bit flips to 1, the $ABD$ bit must automatically flip, too. Consequently, the table above is not part of the actual proof by contraction. 

In contrast, the following rows do make part of the proof:
\begin{equation*}
\begin{array}{l||c|c|c|c|c|c|c|c|c|c}
O & & 0  & 0 & 0 & 0 & 0 & 0 &0 & 0 & \\
    & & \color{Orange}\framebox{1} & \framebox{1} & 0 & 0 & 0 & 0 &0 & 0 & \\
{\rm LHS}~~ & & ~{AB}~ & ~{ABD}~ & ~{ACD}~ & ~{ADE}~ & ~{BCD}~ & ~{BDE}~ & {CEF} & {DEF} & \\
\hline
\hline
{\rm RHS}~~ & ~A~ & B & C & {AD} & {BD} & {DE} & {EF} & ~{ABCD}~ & ~{ABDE}~ & ~{CDEF}~ \\
O & 0 & 0 & 0 & 0 & 0 & 0 & 0 & 0 & 0 & 0 \\
    & ? & ? & ? & ? & ? & ? & ? & \color{Orange}\framebox{1} & \color{Orange}\framebox{1} & ? 
\end{array}
\end{equation*}
After taking into account the LHS nesting relation $AB \subset ABD$, the Hamming distances on both sides of the inequality can be consistently set to 2 by setting all question marks to 0. We conclude that flipping the $AB$-bit away from the $O$-boundary condition does not lead to an unavoidable nesting violation. 

The above discussion illustrates a key difference between nesting violations, which occur between LHS and RHS terms (which are the subject of this paper) versus nesting violations, which occur between LHS terms (which we discard altogether): 
\begin{itemize}
\item Two terms $X(LHS)$ and $Y(RHS)$, which belong on different sides of the inequality, can violate nesting rules~(\ref{nestingrules}) and still be part of a valid proof by contraction. Conditions when this happens are the subject of this paper. 
\item For two terms $X_1 \subset X_2$, which are both on the `greater than' side of the inequality, the implication $x_{X_1} = 1 \Rightarrow x_{X_2} = 1$ is inviolable in the sense that the bit strings which disobey it are discarded from the contraction map altogether \cite{lhsconflicts}. This is because $W(X_1) \cap W(\overline{X_2}) = \emptyset$ so defining the contraction there has no impact for candidate RHS cuts. We eliminated one special case of this complication around (\ref{generalstr}) when we assigned independent bits only to distinct LHS terms $X_i$; this prevented $X_1 = X_2$. But the case $X_1 \subsetneqq X_2$ must be dealt with by hand.
\item For completeness, we remark that a nesting violation between two RHS terms can also be part of a valid proof by contraction. This is why we reserved distinct bits in $f(x)$ for identical RHS terms in equation~(\ref{ffdist}).
\end{itemize}

Before returning to the general discussion, we make one final remark about the example. While flipping the $AB$-bit away from the boundary condition does not produce an unavoidable nesting violation, flipping the $ABD$-bit does:
\begin{equation*}
\begin{array}{l||c|c|c|c|c|c|c|c|c|c}
O & & 0 & 0 & 0 & 0 & 0 & 0 &0 & 0 & \\
    & & 0 & \color{Orange}\framebox{1} & 0 & 0 & 0 & 0 &0 & 0 & \\
{\rm LHS}~~ & & ~{AB}~ & ~{ABD}~ & ~{ACD}~ & ~{ADE}~ & ~{BCD}~ & ~{BDE}~ & {CEF} & {DEF} & \\
\hline
\hline
{\rm RHS}~~ & ~A~ & B & C & {AD} & {BD} & {DE} & {EF} & ~{ABCD}~ & ~{ABDE}~ & ~{CDEF}~ \\
O & 0 & 0 & 0 & 0 & 0 & 0 & 0 & 0 & 0 & 0 \\
    & ? & ? & ? & ? & ? & ? & ? & \color{Orange}\framebox{1} & \color{Orange}\framebox{1} & 0 
\end{array}
\end{equation*}
Here we have a physically realizable pair of bit strings on the LHS side, which are distance 1 away from one another. If we impose nesting rules, the bit strings on the RHS are at least distance 2 away from one another. 

\paragraph{Summary of technique} 

The contrast between flipping the $AB$ bit and the $ABD$ bit reveals how to correctly diagnose unavoidable nesting violations. To inspect the boundary condition for a region $O$, set $O$ to be the purifier and write the inequality in the corresponding canonical form; in this convention $x^O$ and $f^O$ are all zeroes. Now take a LHS term (in the above examples we took $AB$ and $ABD$) and count all terms that contain it, including itself, with coefficients. LHS (`greater-than') terms count with positive coefficients while RHS (`less-than') terms count with negative coefficients. If the total coefficient is negative, a nesting violation is unavoidable. 

We summarize the rule of thumb:
\begin{itemize}
\item Assume that the inequality is presented in the canonical balanced form, which corresponds to setting an atomic region $A$ to be the purifier. 
\item Consider a term $S_X$ on the `greater-than' side of the inequality and compute the total coefficient of all terms that contain $X$, including $X$ itself. 
\item If the total coefficient is negative then imposing nesting rules (\ref{nestingrules}) on a contraction map necessarily leads to a nesting violation. The guaranteed nesting violation occurs between the `all-zeroes' boundary condition for $A$ and the bit string where the $X$-bit is flipped. 
\end{itemize} 

\paragraph{Results} We have verified that the situation described above occurs in \emph{every} boundary condition in 1854 out of the 1868 inequalities being considered. The fourteen exceptions, listed by their labels in \cite{Hernandez-Cuenca:2023iqh}, are inequalities no. 1697 (boundary condition for $A$), nos. 48, 71, 85, 866, 884, 1021 (boundary condition for $B$), nos. 68, 90, 874, 1045, 1155, 1179 (boundary condition for $C$), and no. 106 (boundary condition for $O$). 

A Mathematica notebook, which verifies these claims, is included with the arXiv submission of this paper.

\subsection{Unknown holographic inequalities}
\label{sec:unknown}
We do not know if the statement `contractions, which prove maximally tight holographic inequalities contain nesting violations' admits other exceptions, besides the monogamy of mutual information. Proving this statement rigorously without some essential new insight about the holographic entropy cone seems difficult. Nevertheless, we believe that the statement is true. 

Let us briefly justify this conjecture. Qualitatively speaking, we have found that boundary conditions and nesting rules tend to work against one another in a contraction map. Of course boundary conditions always win because they are unalterable, and this leads to nesting violations. We expect that this characterization applies to all maximally tight holographic inequalities, including the unknown ones. 

A second qualitative finding in this section is that nesting violations in proofs by contraction are not few and far between. They are commonplace, even ubiquitous. In the two infinite families of holographic inequalities, this is especially pronounced near boundary conditions, which seem designed to maximize the density of nesting violations on the attendant tessellations. Is there a rationale behind this? 

We believe that the abundance of nesting violations is related to the maximally tight character (`facetness' for facets of the holographic entropy cone) of the inequalities. To explain this assertion, we invite the reader to imagine a family of nested entropy cones $EC_k$, which are bounded by inequalities provable by contractions with at most $k$ nesting violations:
\begin{equation}
EC_0 \supset EC_1 \supset EC_2 \supset \ldots
\label{ecks}
\end{equation}
From the viewpoint of the true holographic entropy cone $EC = \lim_{k \to \infty} EC_k$, the lower $EC_k$s are bounded by inequalities, which fail to be maximally tight. In this setup, an increasing number of nesting violations $k$ parametrizes the approach to facetness. Of course, chains such as (\ref{ecks}) can formally be defined with respect to any parameters, including arbitrary and useless ones. However, the abundance of nesting violations near boundary conditions of known inequalities appears to be a telling sign. It suggests that counting nesting violations is a useful parameter for quantifying how far an inequality is from being maximally tight (from being a facet of the cone). We return to this point in the Discussion.

\subsection{Nesting violations concern non-empty bulk regions}
Nesting violations in proofs by contraction are puzzling because they make it impossible to interpret the contraction in bulk terms. Ostensibly, the contraction gives us candidate minimal cuts for RHS terms, but those cuts are unphysical from the get-go. There is one conceivable way, however, in which the contraction could retain its putative significance. That would be the case if the bulk regions on which the contraction develops a nesting violation were somehow guaranteed to be empty. (This is the reason why bits strings with $x_{X_1} = 1$, $x_{X_2} = 0$, $X_1 \subset X_2$ are discarded from the contraction.) 

This turns out not to be the case. In other words, nesting violations in proofs by contraction concern bulk regions, which are generically nonempty.

\paragraph{Example} Take inequality (\ref{i5}) from Section~\ref{sec:example}, which we copy below for convenience: 
\begin{align} 
   S_{A_1 A_2 B_1} \,+\, S_{A_2 A_3 B_1}  \,+\, S_{A_3 A_1 B_1} 
& \phantom{~~\geq~~ +.}\!
     S_{A_3 B_1} \,+\, S_{A_1 B_1} \,+\, S_{A_2 B_1} 
\nonumber \\
    +\phantom{.} S_{A_1 A_2 B_2} \,+\, S_{A_2 A_3 B_2} \,+\, S_{A_3 A_1 B_2} 
&  ~~\geq~~  +\phantom{.}\! 
     S_{A_3 B_2} \,+\, S_{A_1 B_2} \,+\, S_{A_2 B_2}  ~~~+\, S_ {A_1 A_2 A_3}  
\nonumber \\    
     +\phantom{.} S_{A_1 A_2 B_3} \,+\, S_{A_2 A_3 B_3}  \,+\, S_{A_3 A_1 B_3}  
& \phantom{~~\geq~~} +\phantom{.}\! 
          S_{A_3 B_3} \,+\, S_{A_1 B_3} \,+\, S_{A_2 B_3}    \label{i5p}
\end{align}
Equation~(\ref{offa2boundary}) identified a LHS bit string, on which a nesting violation is unavoidable. Of all possible targets to which the contraction map can send this bit string, the one with the minimal number (1) of nesting violations is without loss of generality:
\begin{equation}
\begin{tabular}{|ccc|}
\hline
1 & 1 & 0 \\
1 & 1 & 0 \\
1 & 1 & \color{red}{1} \\
\hline
\end{tabular}
~~\xrightarrow{~~\textrm{contraction}~~}~~
\begin{tabular}{|ccc|c}
\cline{1-3}
0 & 0 & 0 & \\
\cline{4-4}
0 & 0 & \color{red}{1} & \multicolumn{1}{c|}{1} \\
\cline{4-4}
0 & 0 & 1 & \\
\cline{1-3}
\end{tabular}
\label{offa2boundary2}
\end{equation}
As in Section~\ref{sec:example}, here we present the contraction as tableaux, which graphically match (\ref{i5p}). The nesting violation is highlighted in red.

Consider a star graph with four legs labeled $A_1, A_2, A_3, B_3$ whose weights are---in this order---1-2-1-1. Regions $B_1$ and $B_2$ are disentangled. The `bulk' region described by the left side of (\ref{offa2boundary2}) is the central node of the star graph. It is nonempty. 

The nesting violation involves the 1 in the $A_2 B_2$ bit on the right. That is, the contraction proposes that the central node should be part of the entanglement wedge of $A_2 B_2$. It does so even though the entanglement wedge of the non-overlapping region $A_1 A_3 B_3$ contains the central node.

\section{A case study in nesting violations: Toric inequalities}
\label{sec:study}

The role played by nesting violations in proofs by contraction---the fact that they are unavoidable and maximally dense---is best appreciated by inspecting an example. Here we focus on the toric inequalities~(\ref{toricineqs}), which were announced and proved in \cite{Czech:2023xed}. We copy the inequalities for the reader's convenience below:
\begin{equation}
\sum_{i=1}^m \sum_{j=1}^n S_{A_i^+ B_j^-}
\geq 
\sum_{i=1}^m \sum_{j=1}^n S_{A_i^- B_j^-} \,+ S_{A_1A_2\ldots A_m}
\tag{\ref{toricineqs}}
\end{equation}
Throughout Section~\ref{sec:study}, in every context where a choice of convention might matter, we take the toric inequalities to be presented in form~(\ref{toricineqs}).

We discuss the contraction map that proves (\ref{toricineqs}) with reference to the tessellation in Figure~\ref{fig:lattices}. LHS terms $S_{A_i^+ B_j^-}$ are associated with faces of the tessellation. RHS terms comprise $mn$ terms $S_{A_i^- B_j^-}$ (which are associated with the vertices of the tessellation) and the special $(mn+1)^{\rm th}$ term $S_{A_1A_2\ldots A_m}$, which is not represented on the tessellation.

\subsection{Proof by contraction of toric inequalities---a review}
\label{sec:toricproofreview}
The requisite contraction map $f$ sends bit strings $x \in \{0,1\}^{mn}$ (which are indexed by LHS terms) to bit strings $f(x) \in \{0,1\}^{mn+1}$ (indexed by RHS terms). On the tessellation, $f$ takes as input a configuration of 0's and 1's that live on the faces of the tessellation, then produces as output an arrangement of 0's and 1's on the vertices of the tessellation, plus an additional bit associated with $S_{A_1A_2\ldots A_m}$. 

\paragraph{Preliminaries}
Reference~\cite{Czech:2023xed} constructed a contraction map using the following ingredients:
\begin{enumerate}
\item Given $x \in \{0,1\}^{mn}$, draw on every face of the tessellation two horizontal or two vertical line segments, depending on whether that face is assigned 0 or 1 in $x$. This is shown in the left panel of Figure~\ref{fig:gamma}.

\begin{figure}[t]
    \centering
    \raisebox{0mm}{
    $\begin{array}{c}
    \includegraphics[width=0.1\linewidth]{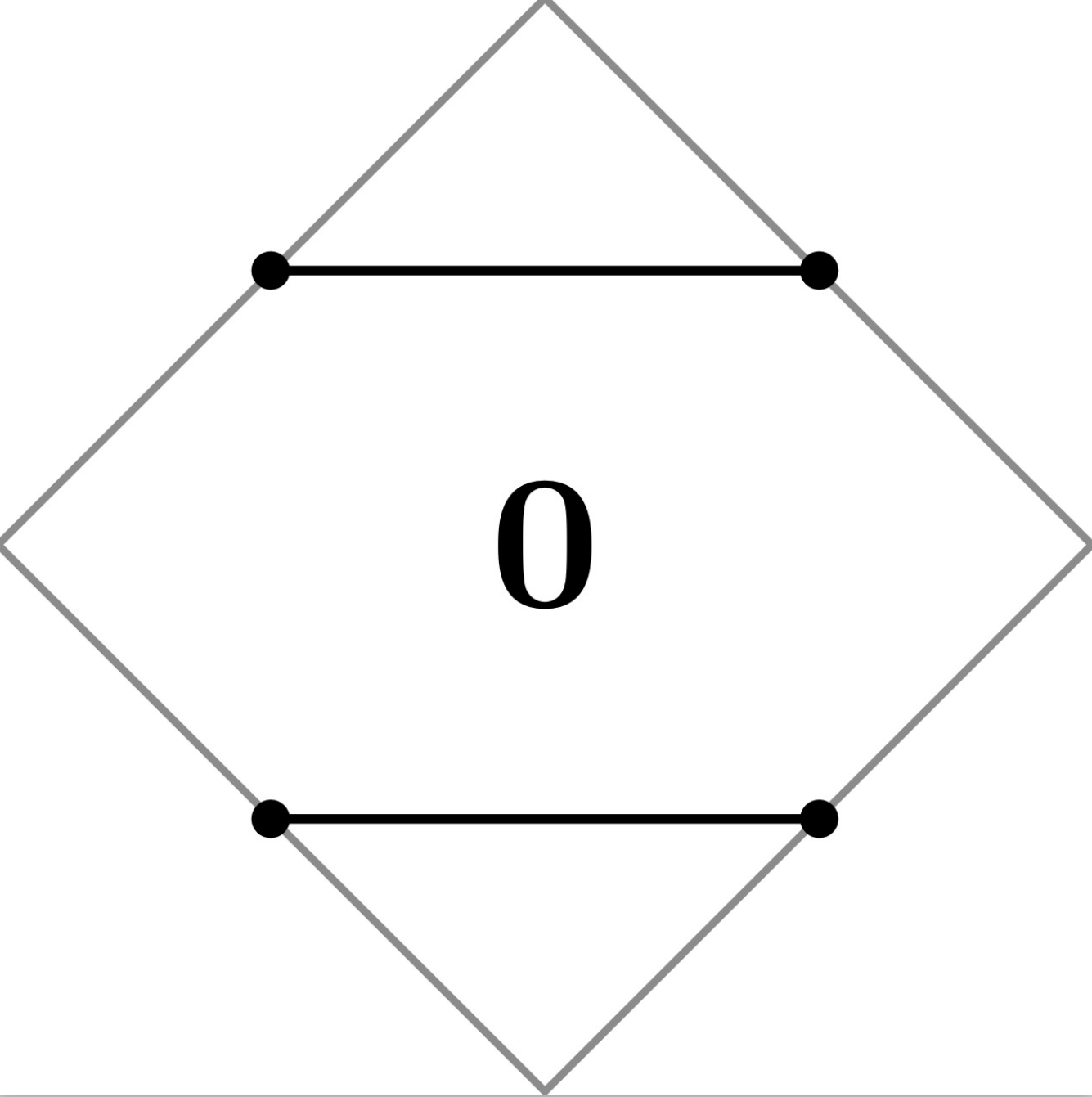} \\
    \includegraphics[width=0.1\linewidth]{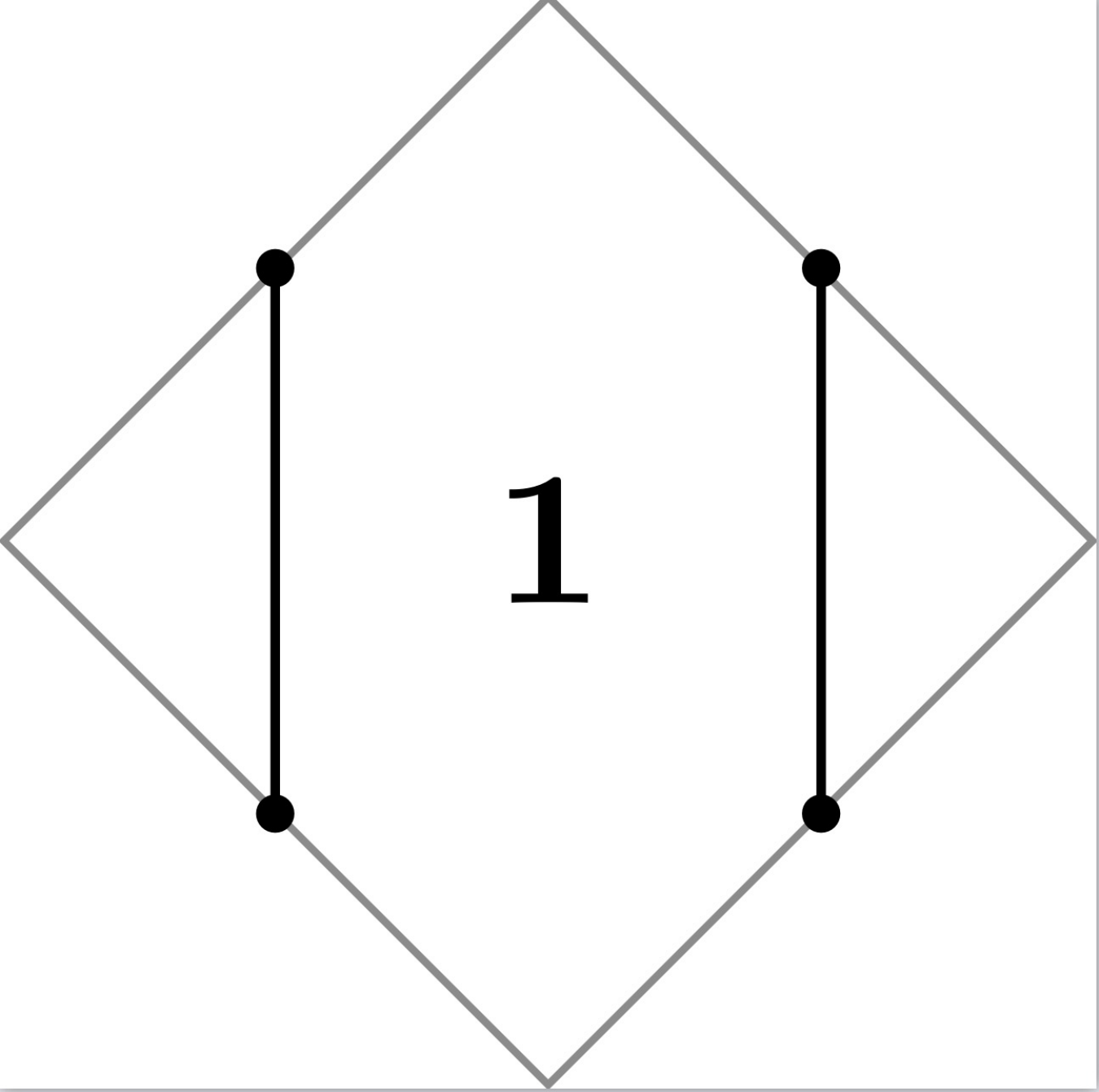}
    \end{array}$}\qquad
    $\begin{array}{cc}
    \includegraphics[width=0.4\linewidth]{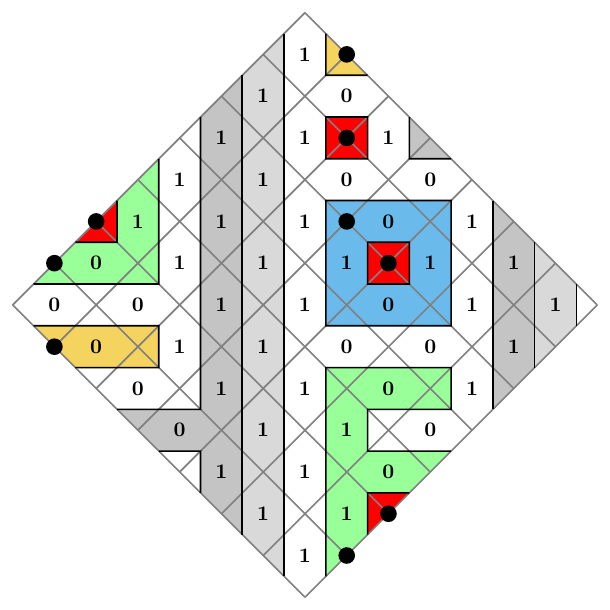} &
    \includegraphics[width=0.4\linewidth]{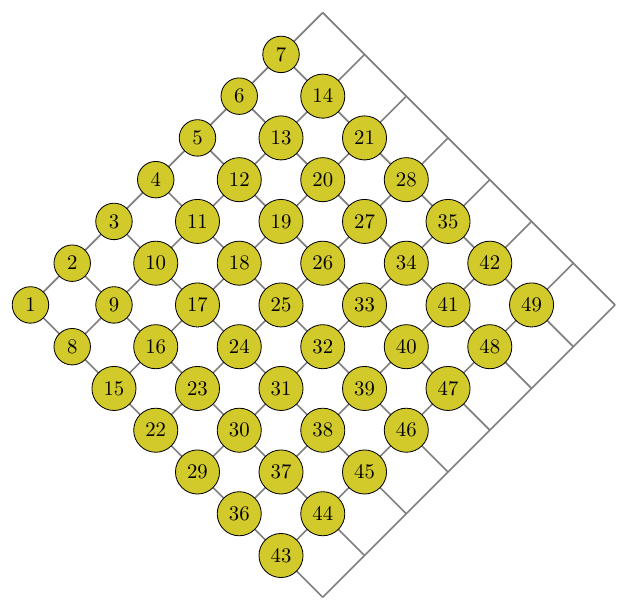}     
    \end{array}$
    
    \caption{ (1) Left panel: The graphical rules for 0's or 1's on faces. (2) Middle panel: An example $\Gamma(x)$. Wrapped connected domains are shaded grey; unwrapped connected domains are in color; minimal unwrapped domains are all {\color{red} red}. (3) Right panel: A ranking scheme for the vertices (lowest number = highest ranking). This ranking scheme is consistent with the middle panel, where the vertices set to 1 are highlighted with \textbullet.}
    \label{fig:gamma}
\end{figure}

The line segments assemble an {\bf $x$-dependent auxiliary graph $\Gamma(x)$}. By construction, $\Gamma(x)$ is a union of disjoint and non-intersecting {\bf loops}, which live on the torus. As such, $\Gamma(x)$ divides the torus into {\bf connected domains $D_k$}. The domains are highlighted in the middle panel of Figure~\ref{fig:gamma}.

It will be useful to distinguish connected domains, which do / do not contain a non-trivial cycle of the torus. We will refer to connected domains, which do not contain a non-trivial cycle as {\bf unwrapped}. Note that an unwrapped connected domain is topologically a disk with any number of punctures. (Examples of unwrapped domains with punctures include the blue and the green domain in the middle panel of Figure~\ref{fig:gamma}.) The boundary of an unwrapped connected domain is a collection of loops in $\Gamma(x)$, each of which wraps the $(0,0)$ cycle on the torus. For brevity, we call loops with wrapping numbers $(0,0)$ {\bf trivial loops}.

\item List all vertices in an arbitrary but definite sequence $\{v_i\}_{i = 1}^{mn}$. The indices in this sequence set up a ranking scheme on the vertices. The ranking scheme will be used to decide which vertex is assigned 1 under $f(x)$, in cases where this choice is not fully determined by other considerations. An example ranking scheme is shown in the right panel in Figure~\ref{fig:gamma}. 

Given a subset of vertices $\{v_{i_1}, v_{i_2} \ldots v_{i_s}\}$, we will refer to the vertex with the smallest index ($v_{i_q}$ such that $i_q = \min\{i_1, i_2 \ldots i_s\}$) as the {\bf highest-ranked vertex} in that subset. 
\end{enumerate}
In summary, the definition of the contraction $f(x)$ uses two ingredients: the $x$-dependent connected domains $D_k$ into which the graph $\Gamma(x)$ divides the torus, and the ranking system $\{v_i\}_{i = 1}^{mn}$ on the vertices.

\paragraph{Definition of contraction \cite{Czech:2023xed}} The bit that corresponds to vertex $v_i$ is set to 1 under $f(x)$ if and only if $v_i$ satisfies the following conditions: 
\begin{itemize}
\item[(1)] The connected domain that includes $v_i$ is unwrapped.
\item[(2)] Vertex $v_i$ is the highest-ranked vertex in the connected domain that contains it.
\end{itemize}
The bits of all other vertices are set to 0.

The $(mn+1)^{\rm th}$ bit---the one associated with the term $S_{A_1 A_2 \ldots A_m}$, which does not correspond to a vertex in the tessellation---is set by:
\begin{equation}
f(x)_{mn+1} = |x| - \#\{\textrm{loops in $\Gamma(x)$}\} + 1 \quad \textrm{(mod 2)}
\label{nongeom}
\end{equation}
Here $|x|$ (mod 2) simply checks whether $x$ contains an even or an odd number of entries set to unity; see equation~(\ref{xxdist}). 

\begin{figure}[t]
		\centering
		\includegraphics[width=0.72\linewidth]{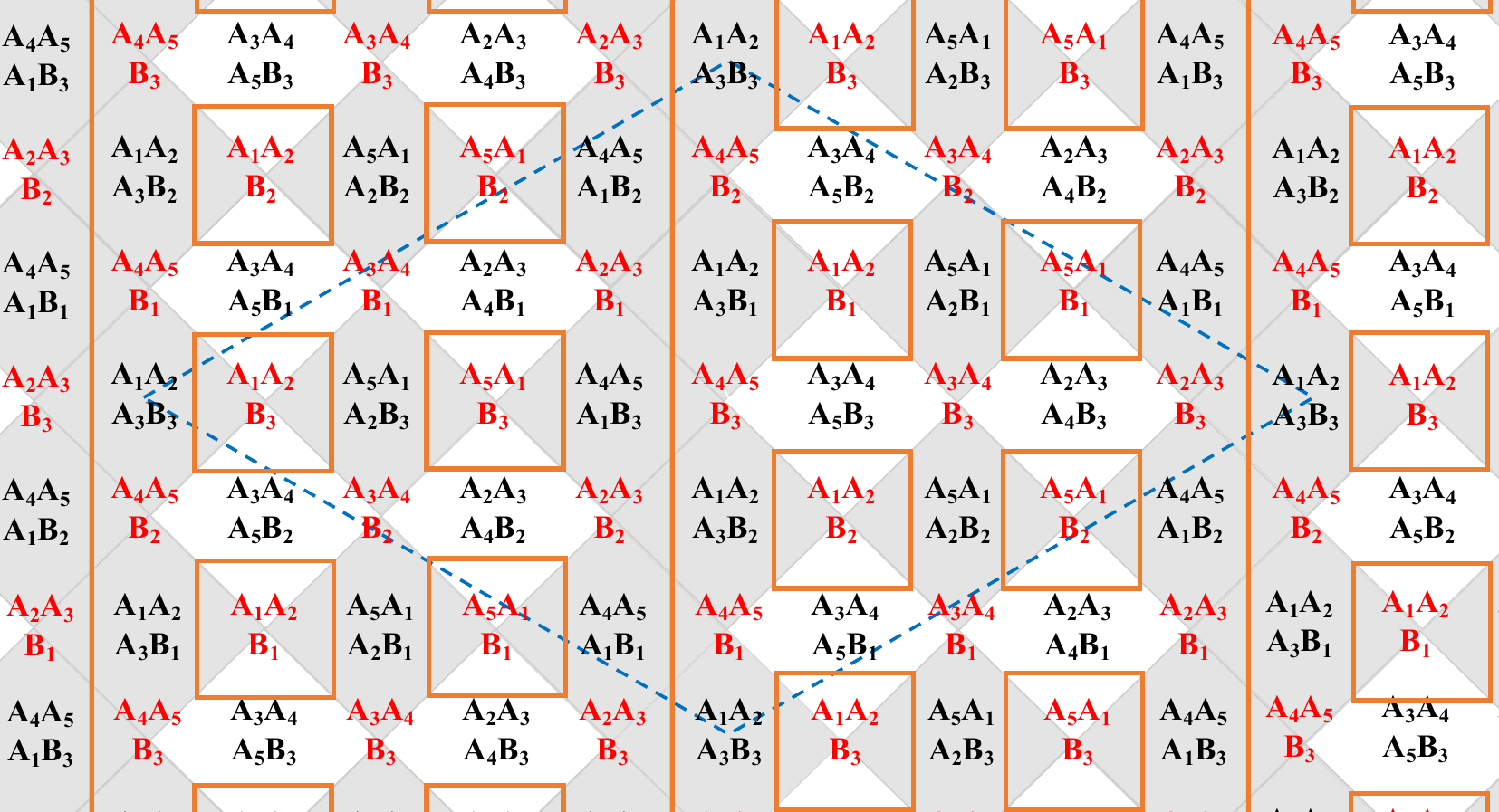} \\
		\vspace*{6mm}
		\includegraphics[width=0.72\linewidth]{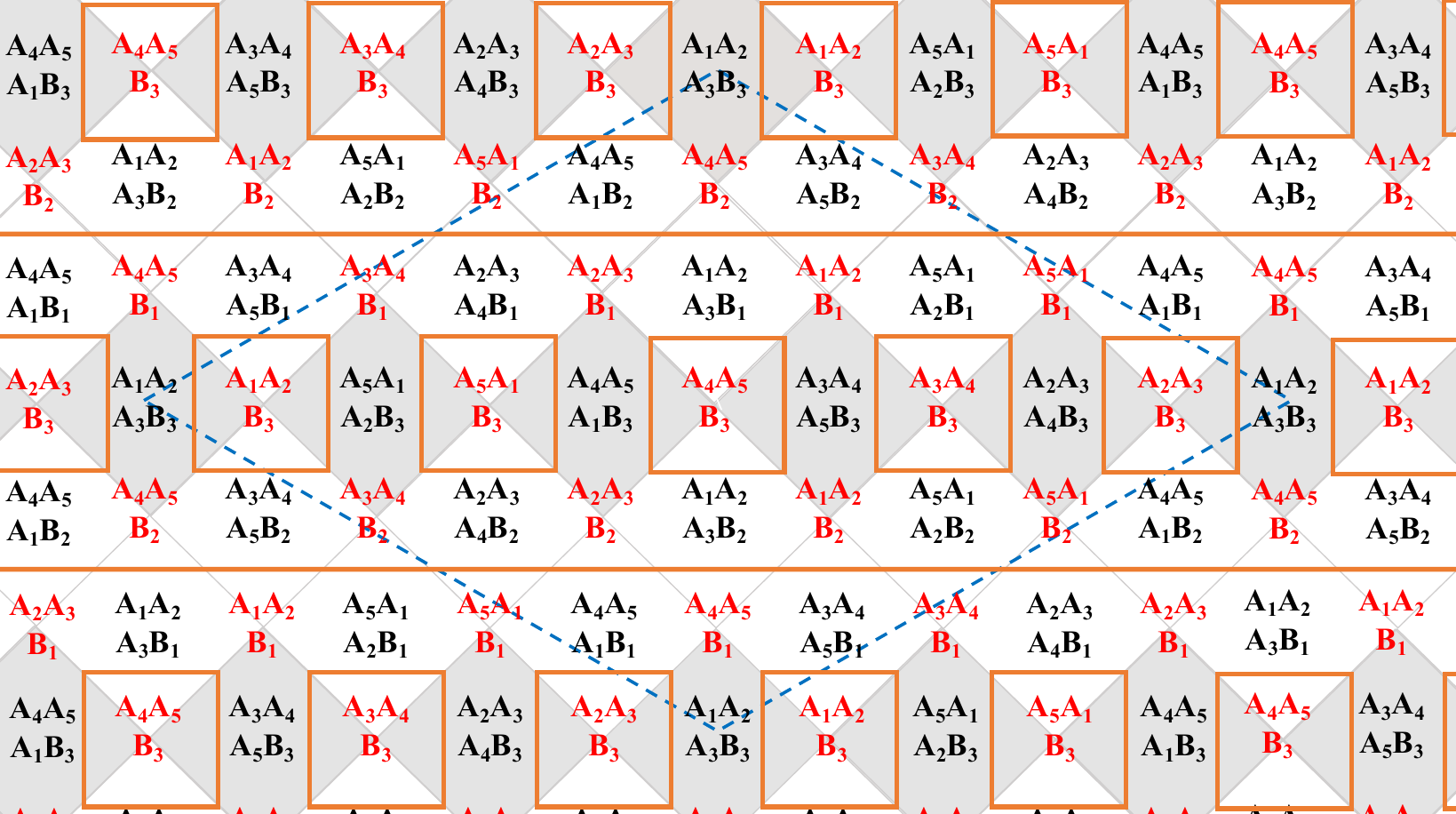} \\
		\vspace*{6mm}
		\includegraphics[width=0.72\linewidth]{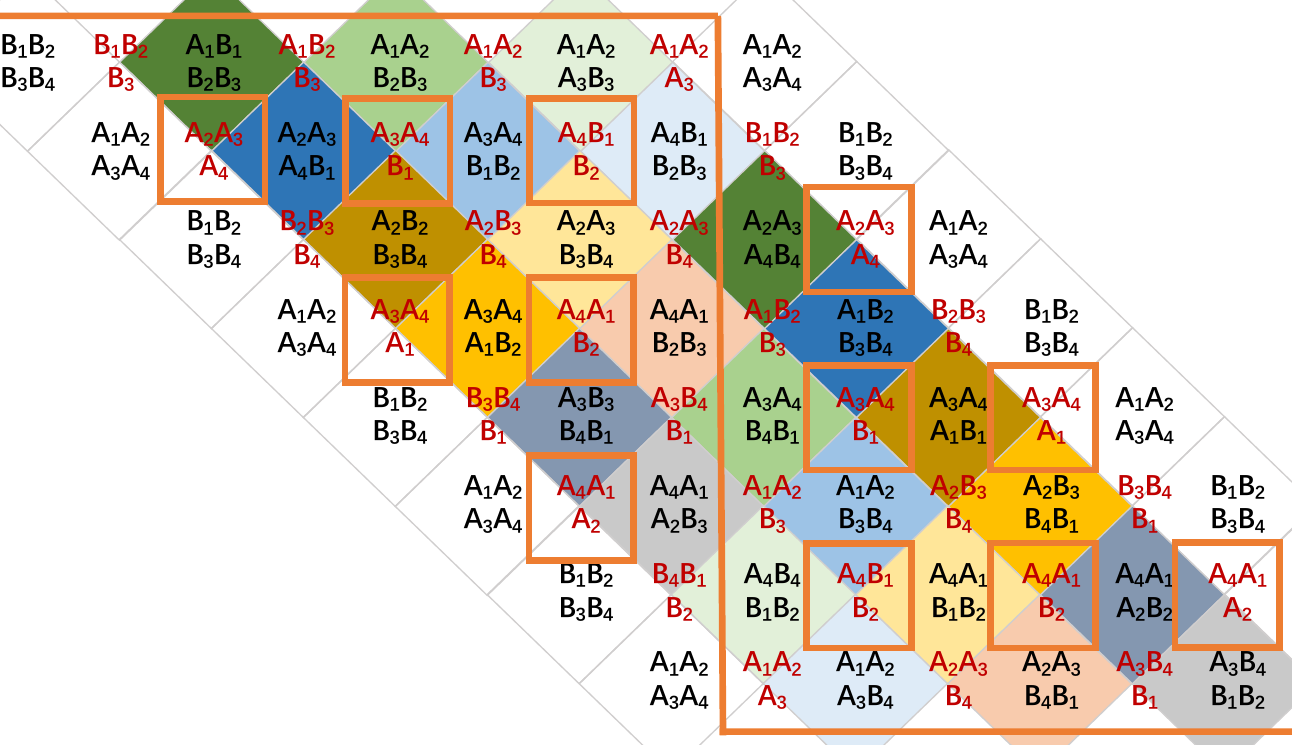}
	\caption{Graphs $\Gamma(x)$, which represent boundary conditions for ${A_1}$ and ${B_3}$ in the $(m,n) = (5,3)$ toric inequality and for $A_4$ in the $m=4$ projective plane inequality. In the upper and middle panels shaded faces have their bits set to 1 in $x$.} 
\label{fig:bcs}
\end{figure}

\paragraph{This $f(x)$ proves the toric inequalities} It is easy to see that map $f(x)$ defined above satisfies the requisite boundary conditions. This is illustrated in Figure~\ref{fig:bcs}. For future use, we remark that every boundary condition contains precisely one nontrivial loop with wrapping numbers $(1, \pm 1)$. The sign tells us whether the boundary condition is associated with an $A$-region or with a $B$-region.

Confirming that $f(x)$ is indeed a contraction requires some rather intricate reasoning. We summarize the argument in Figure~\ref{fig:flips} and develop its physical interpretation in Section~\ref{sec:howtoric}. For a detailed and rigorous proof of the contraction property of $f$ we refer the reader to \cite{Czech:2023xed}.

\subsection{Demystifying the contraction}
\label{sec:howtoric}
As originally laid out in \cite{Czech:2023xed}, the rules above appear rather {\it ad hoc}. In fact, they follow directly from analysing when nesting violations are necessary and when they are avoidable.

\subsubsection{Toric contraction maps are maximally respectful of nesting}
\label{sec:respect}
We show that the contraction maps defined in Section~\ref{sec:toricproofreview} are maximally respectful of nesting. In other words, they violate nesting only when the violation is unavoidable. We develop this claim below.

\paragraph{Minimal loops in $\Gamma(x)$ surround vertices unconstrained by EWN} A minimal loop surrounds the smallest possible connected domain---one that contains a single vertex. As a tautology, this unique vertex is the highest-ranked vertex in the connected domain, so condition~(2) above is satisfied. Furthermore, a minimal connected domain is necessarily topologically trivial, so condition~(1) is satisfied. In summary, every vertex that is surrounded by a minimal loop is automatically set to 1 under $f(x)$. 

Minimal loops in $\Gamma(x)$ have a crisp physical interpretation: They identify those vertices, which can be set to 1 under $f(x)$ without nesting violations. Because all vertices can be set to 0 without nesting violations, this is equivalent to saying that {\bf only vertices surrounded by minimal loops are unconstrained by nesting.} This interpretation of minimal loops is evident from applying the graphical rules for $\Gamma(x)$ (the left panel of Figure~\ref{fig:gamma}) to the bottom-left panel of Figure~\ref{fig:implications}. Doing so draws a minimal loop. 

This discussion reveals that our $f(x)$ is, in a sense, {\bf greedy}. It assigns 1 to {\it every} vertex for which entanglement wedge nesting allows it. However, due to the unavoidable nesting violations that we discussed in Section~\ref{sec:violation}, even this greedy rule is not enough. The contraction needs all the 1's that are allowed by entanglement wedge nesting, {\it and more.}

\paragraph{Two neighboring minimal loops generate an unavoidable nesting violation} Consider a graph $\Gamma(x)$, in which a pair of minimal loops are neighbors. There are two basic scenarios because the pair can be neighbors along the horizontal or the vertical direction. Assuming that both vertices are set to 1 under $f(x)$, these two possibilities are exactly what equations~(\ref{whyviolate1}-\ref{whyviolate2}) describe. 

Recall that equations~(\ref{whyviolate1}-\ref{whyviolate2}) express sufficient conditions for an unavoidable nesting violation. We conclude that:
\begin{itemize}
\item If graph $\Gamma(x)$ contains two nearest neighbor minimal loops and the vertices contained therein are both set to 1 under a contraction $f(x)$ then $f(x)$ contains a nesting violation. This is automatically the case for all maps $f(x)$, which are greedy in the sense defined in the previous paragraph.
\end{itemize}
We can formulate the same reasoning directly in terms of the loops of $\Gamma(x)$. To say that two minimal loops are neighbors is to say that they are supported by a common face. Flipping the bit on that face will destroy the minimality of both loops. Recall that only minimal loops allow a vertex to carry 1 while maintaining consistency with wedge nesting. Thus, to remain consistent with nesting, upon flipping the common face we would have to simultaneously set two vertices to 0. If those two vertices were set to 1 before the flip, setting them now to 0 will violate the contraction property. We conclude that in this circumstance (two vertices, surrounded by nearest neighbor minimal loops in $\Gamma(x)$, both set to 1) a nesting violation is unavoidable.

In Section~\ref{sec:maxdense} we claimed that the boundary conditions in the toric and $\mathbb{RP}^2$ inequalities contain the maximal density of unavoidable nesting violations on the graph. The above analysis makes this claim quantitatively precise. In both cases, the graphs $\Gamma(x)$ have maximally large stretches of the graph covered by a densely packed array of minimal loops. We previously substantiated the claim about the maximal density of nesting violations in Figures~\ref{fig:bctoric} and \ref{fig:rpbc}. Now Figure~\ref{fig:bcs} displays the same fact using loops of $\Gamma(x)$.

\paragraph{Contraction $f(x)$ minimizes nesting violations} Take $x$ such that $\Gamma(x)$ contains two neighboring minimal loops. Our $f(x)$ assigns 1 to both vertices contained therein. Now change $x \to x'$ by flipping the bit on that face, which is common to the two neighboring minimal loops. According to the definition of $f$ given in Section~\ref{sec:toricproofreview}, in $f(x')$ one of the two vertices will remain set to 1 while the other vertex will be reset to 0. 

We can understand this rule as a local minimization of nesting violations. Indeed, any vertex that remains set to 1 in $f(x')$ produces a nesting violation. We already saw that eliminating nesting violations altogether---setting both vertices to 0---would have precluded $f$ from being a contraction. On the other hand, leaving the two vertices set to 1 in $f(x')$ would produce nesting violations on both vertices. By resetting \emph{one} of the vertices to 0 in $f(x')$, map $f$ does the most that can locally be done to remove nesting violations. 

\begin{figure}[t]
    \centering
    $\begin{array}{cc}
    \includegraphics[width=0.48\linewidth]{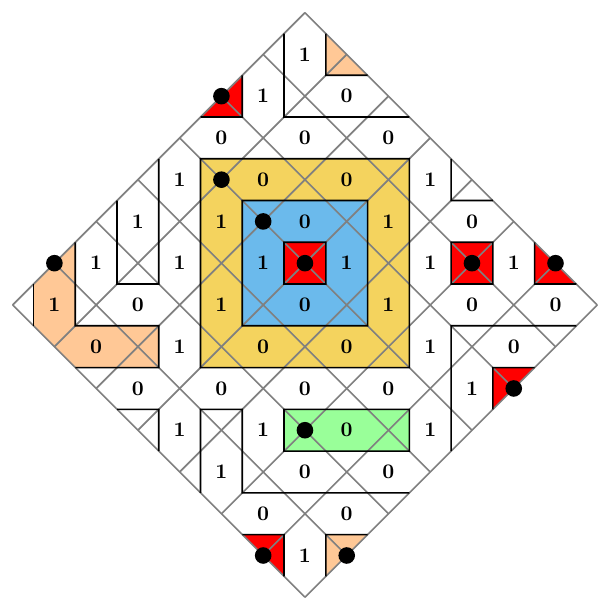} &
    \includegraphics[width=0.48\linewidth]{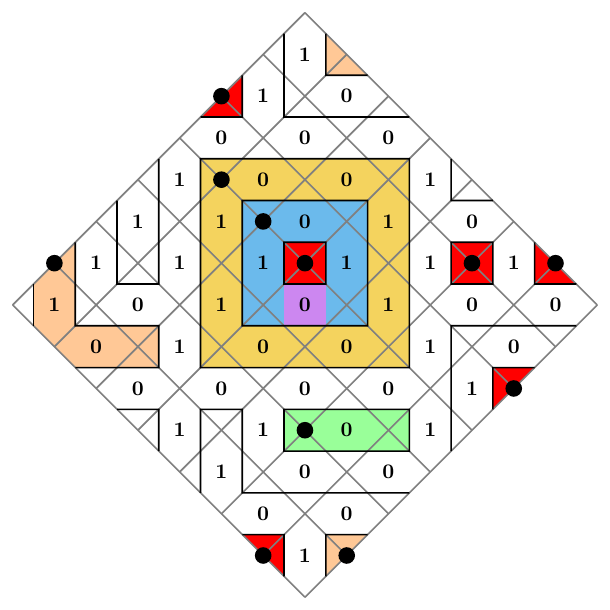} \\
    \includegraphics[width=0.48\linewidth]{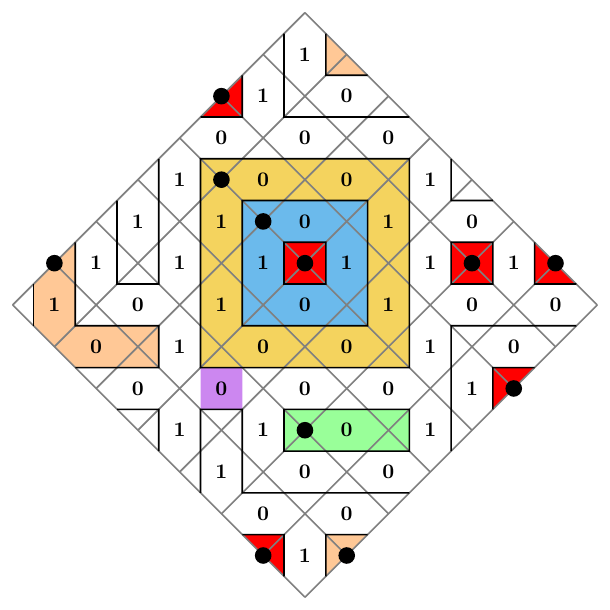} &
     \includegraphics[width=0.48\linewidth]{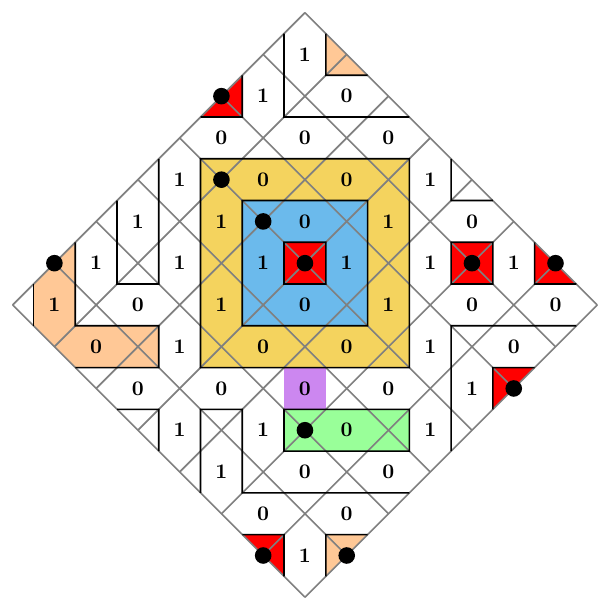}
    \end{array}$
        \caption{Why map $f$ is a contraction. Top left panel: We start with a specific $\Gamma(x)$; unwrapped connected domains are highlighted in color; the vertices mapped to 1 under $f(x)$ are distinguished with \textbullet~; the vertex ranking scheme is the same as in Figure~\ref{fig:gamma}. Consider three bit flips, which act on the faces shown in purple. Bottom left panel: This bit flip merges the yellow unwrapped domain with the white wrapped domain so \emph{one} \textbullet~(from the yellow domain) flips to 0. Right panels: These bit flips merge two unwrapped domains (yellow and red / yellow and green) into one unwrapped domain so \emph{one} \textbullet~(the lower-ranked one, from the red / green domain) flips to 0. \\
$\phantom{~~~~~}$Two other possible effects of a bit flip are not shown here: (1) A bit flip that merges two wrapped loops of $\Gamma(x)$ into one unwrapped loop; this creates \emph{one} \textbullet~; an illustrative example is to flip any one bit in the light grey domain in Figure~\ref{fig:gamma}. (2) A bit flip that re-wires a single loop of $\Gamma(x)$; this is discussed in Figure~\ref{fig:extra_bit} and it flips the $S_{A_1A_2 \ldots A_m}$ bit. In all cases, one face bit flip alters precisely \emph{one} bit of $f(x)$ so map $f$ is a contraction.}
    \label{fig:flips}
\end{figure}

This logic does not rely on the loops being minimal. By induction on the number of vertices in the connected domain, it applies to every bit flip that merges two \emph{unwrapped} domains into one. Indeed, consider two connected domains $D_1$ and $D_2$ demarcated by $\Gamma(x)$, which contain $k_1$ and $k_2$ vertices, respectively. Further, suppose that flipping a single bit in $x \to x'$ changes $\Gamma(x)$ in such a way that $D_1$ and $D_2$ merge into one connected domain; see the right panels in Figure~\ref{fig:flips}. This means that two topologically trivial loops in $\Gamma(x)$ merge into a single unwrapped non-minimal loop in $\Gamma(x')$. By the inductive hypothesis, before the flip $D_1$ and $D_2$ each contain one vertex set to 1 under $f(x)$. After the flip, if either vertex remains set to 1 under $f(x')$, it will be a nesting violation. Map $f$ resets one of them to 0 under $f(x')$ but it cannot reset both. In this way, $f$ does \emph{the most it locally can} to remove nesting violations.

We highlight a pithy consequence of the above reasoning: {\bf Every non-minimal loop of $\Gamma(x)$ encircles one unavoidable nesting violation}. 

\paragraph{Nesting violations in wrapped domains are unnecessary} The contraction $f$ assigns 0 to all vertices contained in wrapped domains. If any vertex in a wrapped domain were set to 1, it would be an avoidable nesting violation. Moreover, when a bit flip adjoins an unwrapped domain to a wrapped domain, the vertex  in the unwrapped domain that carries 1 is immediately reset to 0, thus eliminating a potential nesting violation; see the bottom left panel of Figure~\ref{fig:flips}. Altogether, the fact that all vertices in wrapped connected domains are set to 0 realizes the principle of minimizing nesting violations. 

Let us comment on why map $f$ treats unwrapped and wrapped domains differently. We can `build up' an unwrapped domain $D$ with $k$ vertices in it by starting from $k$ minimal loops in $\Gamma(x_{\rm initial})$ and performing $k-1$ bit flips, adjoining one minimal loop at a time. Because $f$ is a contraction, this sequence can remove at most $k-1$ out of the $k$ 1's in $f(x_{\rm initial})$. In other words, having at least one vertex set to 1 is a topological property of every unwrapped domain. (This implies a nesting violation when the domain is not minimal.) In contrast, merging $k$ minimal loops into a \emph{wrapped} domain requires at least $k$ bit flips: $k-1$ flips to glue up the domain and at least one additional flip to close a topologically nontrivial cycle. As a consequence, \emph{every} vertex in a wrapped domain has an opportunity to be reset to 0.  

\paragraph{Schemes to select nesting violations}  
In a given unwrapped, non-minimal connected domain $D$, the vertex to be set to 1 might a priori be chosen in a context-dependent way. That is, the choice might depend on all of $x$ and reflect properties of graph $\Gamma(x)$ as a whole. Let us make the simplifying assumption that the chosen vertex depends only on domain $D$ and knows nothing about how the torus is partitioned outside of $D$. We call this vertex $v(D)$. To recapitulate: $v(D)$ is the special vertex in the unwrapped domain $D$, which is assigned 1 by $f(x)$ when $D$ is a connected domain wrt $\Gamma(x)$.  

The contraction property of map $f$ imposes a constraint on function $v(D)$. Consider what happens when a bit flip $x \to x'$ merges two unwrapped domains $D_1$ and $D_2$. Then we must have:
\begin{equation}
v(D_1 \cup D_2) = v(D_1) \quad {\rm or} \quad v(D_2)
\label{vdcondition}
\end{equation}
Otherwise, a bit flip that merges $D_1$ with $D_2$ would reset vertices $v(D_1)$ and $v(D_2)$ from 1 to 0 and reset the vertex $v(D_1 \cup D_2)$ from 0 to 1. That would make $|f(x)-f(x')| = 3$ and $f$ would not be a contraction.  For a given unwrapped domain $D$, (\ref{vdcondition}) must hold for every decomposition of $D$ into neighboring domains $D_1$ and $D_2$.

There is more than one way to satisfy~(\ref{vdcondition}). For example:
\begin{itemize}
\item On an orientable surface such as the torus, we can select the vertex that occupies some relative position within $D$. For example, Reference~\cite{Czech:2023xed} set $v(D)$ to be the vertex which is `bottommost and, if there is a tie, rightmost within $D$.'
\item A global ranking system: $v(D) = v_{i_s}$ such that $i_s = \min\{i_q\, |\, v_{i_q} \in D\}$. This approach is used in Section~\ref{sec:toricproofreview}. 
\end{itemize}
It is easy to see that both options satisfy (\ref{vdcondition}). There exist other schemes to define $v(D)$. As mentioned above, it is also possible to choose the vertices to be set to 1 in a `non-local' way, which cannot be described using a single function $v(D)$. 

\paragraph{Choices in $f$ are about nesting violations} Thus far, we have found that a contraction map, which respects nesting as much as possible (makes as few nesting violations as possible) must have the following properties:
\begin{itemize}
\item It assigns 1 to one vertex in each unwrapped domain demarcated by $\Gamma(x)$.
\item It assigns 0 to all vertices in all wrapped domains demarcated by $\Gamma(x)$.
\end{itemize}
If we stick to this structure, the non-uniqueness of $f$ originates from the choice of \emph{which} vertex in an unwrapped, non-minimal domain will be involved in a nesting violation.\footnote{Conceivably, setting the bit for the $(mn+1)^{\rm th}$, global term $S_{A_1 A_2 \ldots A_m}$ might introduce additional ambiguities in $f$. This term is not in a nesting relationship with any other term in the inequality (except in the dihedral case $n=1$), so any extra ambiguities in that bit alone would be unrelated to nesting violations. We think this possibility is unlikely because (1) in the contractions we describe, this bit is a topologically protected quantity (see Section~\ref{sec:global}) and (2) we have generated multiple contraction maps numerically and always found this term to behave rigidly. Nevertheless, we have not rigorously eliminated this possibility. \label{foot:other}} 
Moreover, any departure from the aforementioned structure would gratuitously introduce additional nesting violations. We conclude that all choices and ambiguities in the contraction map perform one function: they select the pattern of nesting violations.

\begin{figure}[t]
    \centering
\includegraphics[width=0.9\linewidth]{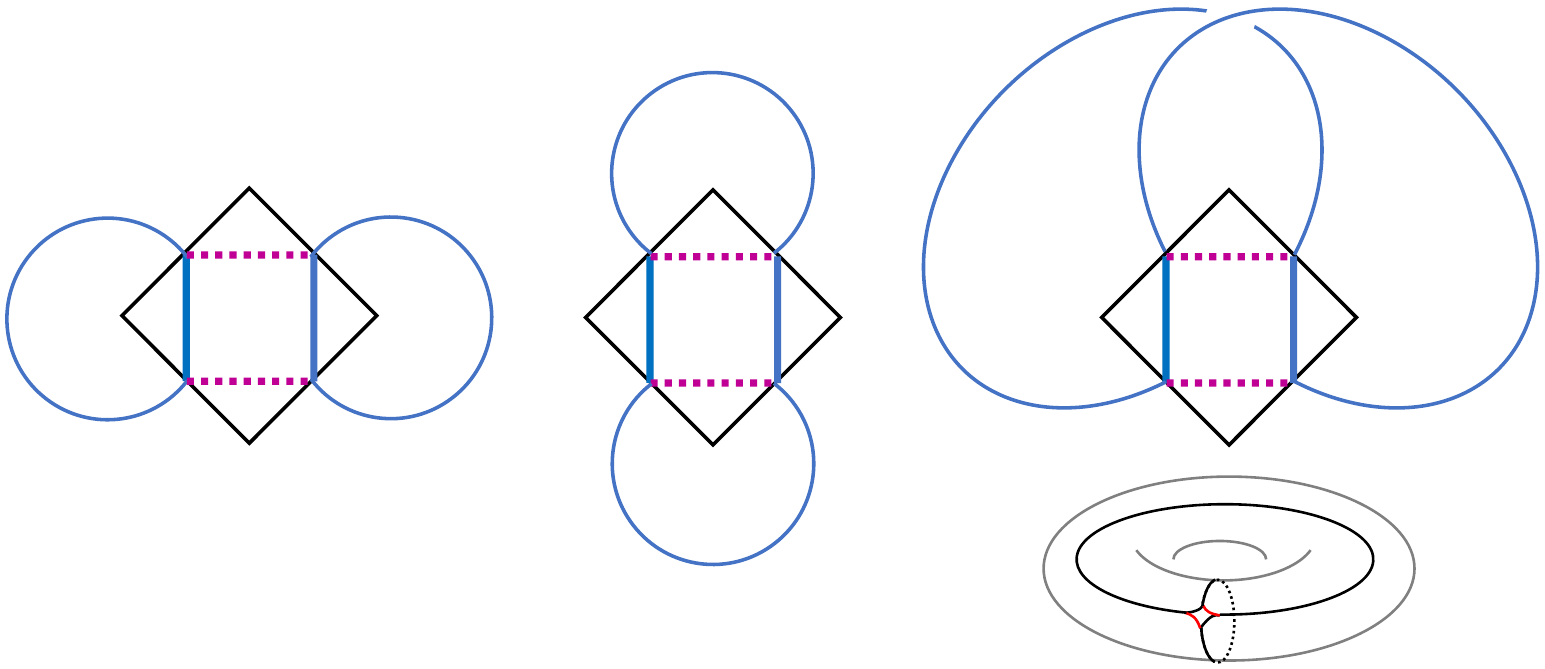}
    \caption{A schematic representation of the loops of $\Gamma(x)$, which pass through a given face. Away from the face, they can be wired in three different ways: connecting left to left and right to right (leftmost panel); top to top and bottom to bottom (middle panel); or across (rightmost panel). In the former two cases, a bit flip merges two loops into one or vice versa. In the latter case, a bit flip rewires a single loop. This case could not arise on the plane because the loops would have to intersect, but it can and does arise on the torus.}
    \label{fig:extra_bit}
\end{figure}

\subsubsection{The non-geometric bit for $S_{A_1 A_2 \ldots A_m}$}
\label{sec:global}
Under most circumstances, a bit flip in $x$ causes two loops in $\Gamma(x)$ to merge into one loop (or vice versa). Such bit flips decrease or increase the number of loops in $\Gamma(x)$ by one. Occasionally, however, flipping one bit in $x$ may leave the number of loops in $\Gamma(x)$ unchanged; see Figure~\ref{fig:extra_bit}. This scenario plays an essential role in setting the $(mn+1)^{\rm th}$ bit in $f(x)$---the one that accounts for the global term $S_{A_1 A_2 \ldots A_m}$ in the toric inequalities.

Let us examine a single face undergoing a bit flip; see Figure~\ref{fig:extra_bit}. Four ends of loops reach this face from elsewhere on the torus. Outside the face undergoing the flip, the four ends join up into two segments. The effect of the bit flip on the loops of $\Gamma(x)$ depends on how they do so. In the cases shown in the left and middle panel of Figure~\ref{fig:extra_bit}, the bit flip causes two distinct loops to merge into one loop or vice versa. But in the case shown in the right panel, the bit flip rewires a single loop.

The next step in our discussion is the following deduction:
\begin{itemize}
\item Premise (justified below): If a bit flip rewires a single loop in $\Gamma(x)$ then that loop is topologically nontrivial on the torus. 
\item Therefore, a bit flip that rewires a single loop in $\Gamma(x)$ has no effect on unwrapped domains. That is because unwrapped domains have boundaries, which are (unions of) trivial loops whereas here we are only modifying a topologically nontrivial loop.
\item Therefore, a bit flip that rewires a single loop in $\Gamma(x)$ does not affect the $m \times n$ bits of $f(x)$, which are associated with the vertices of the tessellation. That is because those bits only depend on the pattern of unwrapped domains.
\item Therefore, bit flips that rewire a single loop can only be relevant to setting the $(mn+1)^{\rm th}$, non-geometric bit in $f(x)$.
\end{itemize}
We now justify the premise of this argument. After that, we discuss how exactly the $(mn+1)^{\rm th}$ bit of $f(x)$ is set.

\paragraph{Only nontrivial loops can be rewired instead of merging or breaking} We review the argument from \cite{Czech:2023xed}. Consider the configuration shown in the right panel in Figure~\ref{fig:extra_bit} but make one modification: instead of drawing a pair of horizontal or vertical lines like we do in $\Gamma(x)$, draw a cross \text{\sffamily X}. This modification produces two distinct loops with intersection number 1; we know they have no intersections away from the \text{\sffamily X} because $\Gamma(x)$ consists of non-intersecting loops. Denoting the wrapping numbers of the two loops with $(p,q)$ and $(p', q')$, we have:
\begin{equation}
\det \left( \begin{array}{cc} p & p' \\ q & q' \end{array} \right) = \pm 1
\label{intcond}
\end{equation}
In terms of $(p,q)$ and $(p', q')$, the loop of $\Gamma(x)$ that is being rewired has wrapping number $(p,q) \pm (p', q')$. The two signs describe the loop before and after the bit flip rewires it.

Equation~(\ref{intcond}) does not allow $(p,q) \pm (p', q') = (0,0)$. This establishes the premise of the preceding argument: If a bit flip in $x$ rewires a loop in $\Gamma(x)$ rather than breaking it into two or merging it with another then that loop is topologically nontrivial on the torus. 

We add a few remarks:
\begin{itemize}
\item The possibility that a bit flip leaves the number of loops in $\Gamma(x)$ unchanged has a topological origin; it is only there because the torus is not simply connected. This is evident from equation~(\ref{intcond}): a bit flip that preserves the number of loops implies the existence of two cycles with intersection number 1.
\item The simplest possibilities for a loop to be rewired are $(1,0) \pm (0,1)$. These are the transitions between loops $(1, \pm 1)$, which appear in boundary conditions; see Figure~\ref{fig:bcs}. But there are other possibilities, for example:
\begin{equation}
(2,1) - (1,0) = (1,1) \quad \longleftrightarrow \quad (2,1) + (1,0) = (3,1)  
\end{equation}
Proofs by contraction of toric inequalities with sufficiently many regions ($m$ and $n$ high enough) generally feature loops with arbitrarily high wrapping numbers.
\end{itemize}
\smallskip

\paragraph{Setting the final bit} The contraction map $f(x)$ defined in Section~\ref{sec:toricproofreview} takes \emph{every} opportunity to flip its final bit. That is, the $(mn+1)^{\rm th}$ bit of $f(x)$ changes value if and only if:
\begin{equation}
\#\{\textrm{loops in $\Gamma(x)$}\} = \#\{\textrm{loops in $\Gamma(x')$}\}
\tag{reset the final bit}
\end{equation} 
As stated, this rule may appear path-dependent. That is, if $(x,x')$ and $(x,x'')$ both differ in a single bit, can we be sure that applying the rule will set the final bit of $f(x)$ consistently? This question is answered by equation~(\ref{nongeom}):
\begin{equation}
f(x)_{mn+1} = |x| - \#\{\textrm{loops in $\Gamma(x)$}\} + 1 \quad \textrm{(mod 2)}
\tag{\ref{nongeom}}
\end{equation}
It is clear that (\ref{nongeom}) is both well-defined and consistent with the rule.

Conceivably, there might exist contraction maps, which are maximally respectful of nesting (in the sense defined in Section~\ref{sec:respect}) yet do not obey (\ref{nongeom}); see also footnote~\ref{foot:other}. But we have never encountered such a contraction map, despite having generated numerically hundreds of contraction maps using \cite{contractor}. 

\paragraph{Significance} We offer the following remarks about the final bit of $f(x)$.
\begin{itemize}
\item This bit is the raison d'{\^e}tre of the toric inequalities. Without the final term on the RHS, the toric inequalities would be combinations of strong subadditivities; viz. discussion around inequality~(\ref{cmirewrite}).
\item As we remarked before, flips that preserve the number of loops in $\Gamma(x)$ are only possible because of the toroidal topology of the graph model.  
\item Therefore, the improvement on strong subadditivity offered by the toric inequalities is truly topological in character. It arises \emph{because} the nesting relations among regions $A_i^+ B_j^{\pm}$ make up a structure, which is not simply connected. 
\end{itemize}

We should view equation~(\ref{nongeom}) as defining a $\mathbb{Z}_2$-valued topological index on $x \in \{0,1\}^{mn}$, understood as bits living on faces of a toroidal tessellation. In other words, it divides up $x \in \{0,1\}^{mn}$ into two classes: those with $f(x)_{mn+1} = 0$ and those with  $f(x)_{mn+1} = 1$. 

Stated in terms of candidate RT surfaces, the intended meaning of this index is clear: it attaches the bulk region $W(x)$ to the candidate entanglement wedge of $A_1 A_2 \ldots A_m$ or to the candidate entanglement wedge of $B_1 B_2 \ldots B_n$. However, because candidate RT surfaces assembled according to contraction maps do not respect nesting, this interpretation of (\ref{nongeom}) should be taken with a grain of salt. We therefore posit a question: Can we give quantity (\ref{nongeom}) a different, more physical interpretation? 

\subsubsection{Summary} 
\label{sec:toricsummary}
We have found that contractions maps with which \cite{Czech:2023xed} proved the toric inequalities have the following properties:
\begin{itemize}
\item Demanding that the contraction commit as few nesting violations as possible fixes the structure of the maps given in \cite{Czech:2023xed}, up to their essentially non-unique features.
\item The non-unique features of the contractions reflect choices for how the contraction violates nesting.
\item The toric inequalities improve on strong subadditivity for a reason, which is topological in character. The topology in question concerns the nesting relations among terms in the inequality. 
\end{itemize}

\subsection{Proofs by contraction of the projective plane inequalities}
\label{sec:howrp2}

The projective plane inequalities take the form:
\begin{equation} 
\!\!\!
\frac{1}{2} 
\sum_{j=1}^{m-1} \sum_{i=1}^{m}
\left(S_{A_i^{(j)} B_{i+j-1}^{(m-j)}} + S_{A_i^{(j)} B_{i+j}^{(m-j)}} \right)
\,+\, (m-1)\,S_{A_1 A_2 \ldots A_m}
\geq
\sum_{i,j=1}^{m} S_{A_i^{(j-1)} B_{i+j-1}^{(m-j)}}
\tag{\ref{rp2ineqs}}
\end{equation}  
As explained in Section~\ref{sec:rp}, the terms in the first sum come in complementary pairs so in fact they all appear with coefficient 1. We further saw in Section~\ref{sec:rp} that these inequalities too can be described by a graphical model. It features a square tessellation of a M{\"o}bius strip glued to a single $(2m)$-gon, which corresponds to the special term $(m-1)\,S_{A_1 A_2 \ldots A_m}$. This almost-regular tiling of the projective plane has $m^2 - m$ squares and one $(2m)$-gon, so altogether the bit strings $x$ live in $\{0,1\}^{m^2 - m + 1}$. 

Reference~\cite{Czech:2023xed} proved the projective plane inequalities by contraction. It turns out that the results of the previous subsection---structural properties of the toric contraction maps, which are explained by entanglement wedge nesting---carry over unchanged to proofs of~(\ref{rp2ineqs}). A detailed analysis of the relevant contractions vis-{\`a}-vis entanglement wedge nesting would read almost identical to Section~\ref{sec:respect}.

To be concrete, we list the salient common points:
\begin{itemize}
\item We define a graph $\Gamma(x)$ and divide the projective plane into connected domains. We again distinguish wrapped and unwrapped domains.
\item For a given $x \in \{0,1\}^{m^2-m+1}$, the minimal loops in $\Gamma(x)$ surround those vertices, to which $f(x)$ can assign 1 or 0 without a nesting violation. Neighboring unwrapped loops indicate unavoidable nesting violations. Configurations that induce unavoidable nesting violations occur in boundary conditions with maximal density; see Figure~\ref{fig:bcs}. Every non-minimal unwrapped loop encircles a vertex whose bit in $f(x)$ violates nesting.
\item Map $f(x)$ assigns 1 to the highest-ranking vertex in every unwrapped domain and 0 to all other vertices. This $f$ again has the property that it minimizes the number of nesting violations whenever a bit flip in $x$ merges two loops in $\Gamma(x)$ into one. The argument why this is so is identical to that in Section~\ref{sec:respect}.
\item The ambiguity in defining $f(x)$ is again about choosing which vertex is assigned 1 in non-minimal unwrapped domains. This chooses \emph{which} unavoidable nesting violations $f$ commits.  
\end{itemize}

We also briefly list the differences:
\begin{itemize}
\item The projective plane is not orientable. Therefore, choosing the nesting violation according to a relative position in the domain (`bottommost and, if there is a tie, rightmost within $D$' in toric inequalities) is not available. Reference~\cite{Czech:2023xed} used the ranking scheme.
\item There is no analogue of the $(mn + 1)^{\rm th}$ bit of the toric inequalities. Every bit flip in $x$ merges two loops of $\Gamma(x)$ into one or vice versa.
\item The only thing, which is more complex in the projective plane inequalities is the analysis of what happens when we flip the bit in $x$, which corresponds to the special term $(m-1)\,S_{A_1 A_2 \ldots A_m}$. To preserve the contracting character of map $f$, flipping that bit in $x$ can cause at most $m-1$ bits in $f(x)$ to flip. Establishing that this condition is not violated requires a somewhat intricate argument, which combines theorems in topology with the theory of non-crossing permutations. For details, see \cite{Czech:2023xed}.
\end{itemize}

\section{Discussion}
\label{sec:discussion}

The main finding of this paper is that contraction maps used in proving holographic entropy inequalities define candidate RT surfaces, which are not consistent with entanglement wedge nesting. Violations of nesting in proofs by contraction are not sporadic exceptions. They occur routinely and---in cases where a graph model can quantify their incidence---with maximal possible density. 

The most straightforward lesson, which one might draw from these findings is that the content of contraction maps should not be interpreted in a physical way. In this view, contractions are mathematical tools to establish the inequalities and nothing else. However, if that perspective were correct and complete, one would not expect considerations of nesting to play any role in constructing and analysing contractions. 

Our study of contraction maps, which prove the toric inequalities (\ref{toricineqs}) and the projective plane inequalities (\ref{rp2ineqs}) indicates otherwise. We found that the rather complicated structure of these contraction maps follows in its entirety from stipulating that the contraction should be `as consistent with nesting as possible.' A whole catalogue of auxiliary concepts---tessellations of two-dimensional manifolds, graphs $\Gamma(x)$ and their component loops, wrapped and unwrapped domains---turn out to characterize aspects of entanglement wedge nesting. A particularly intriguing feature of the analysis is the central importance of the topology of $T^2$ and $\mathbb{RP}^2$ for constructing and understanding the contractions. As we saw in rewritings~(\ref{cmirewrite}) and (\ref{rp2ineqs4}), rules based on entanglement wedge nesting patch these manifolds together from conditional mutual informations in a manner, which is reminiscent of the construction of kinematic space \cite{Czech:2015qta}.

It would be fascinating to find a direct physical interpretation of the contractions maps, which is not subordinate to holographic entropy inequalities. Because the mechanism that makes holographic entropy inequalities true is phase transitions in RT surfaces---and because the modern way of understanding such phase transitions involves the erasure correcting properties of the bulk \cite{Almheiri:2014lwa}---we anticipate that the primary area of application of contraction maps is holographic error correcting codes \cite{Pastawski:2015qua}. One specific way in which EWN violations in contraction maps constrain holographic error correction will be discussed in a forthcoming paper \cite{violationHEC}. There are also striking similarities to other aspects of physics. Another forthcoming paper \cite{2sidedDE} will discuss a relationship with the differential entropy formula \cite{Balasubramanian:2013lsa,Czech:2014tva,Czech:2015qta}.

Readers who are sceptical about directly interpreting contraction maps might be motivated by our findings to pursue another goal: To find an alternative method to prove holographic entropy inequalities. If the logic of the contraction method is to produce candidate RT surfaces, and if the candidate RT surfaces are guaranteed to be an unphysical fantasy, it would be surprising if Nature and Geometry did not equip us with a less chimerical way of establishing the inequalities. The notorious difficulty in extending proofs by contraction to time-dependent settings \cite{Czech:2019lps, mattnew} where the HRT proposal replaces RT is perhaps another indication that we should change tactics. In this paper, we further argued that most (perhaps all) of the ambiguities and choices in proofs by contraction concern precisely that part of the contraction table, which is inconsistent with wedge nesting and therefore, in bulk terms, unphysical. We posit a challenge: How to prove holographic entropy inequalities without engineering an elaborate falsehood?

Our analysis sheds some light on the structure and meaning of holographic entropy inequalities. In the toric and projective plane inequalities, we found that nesting violations occur in the neighborhoods of boundary conditions with maximal density. In other known holographic entropy inequalities, we found that a nesting violation is present in the neighborhoods of nearly all individual boundary conditions, with only a handful of exceptions. However, once the boundary conditions were set, we saw that the known contraction maps for the toric and projective plane inequalities were structured so as to minimize nesting violations. 

Based on these observations, one may speculate that maximally tight holographic entropy inequalities and their contraction proofs are solutions of a {\bf maximin} condition on nesting violations:
\begin{itemize}
\item For fixed boundary conditions, find a contraction map $f$ that commits as few nesting violations as possible.
\item Set boundary conditions so that contraction $f$ commits as many nesting violations as possible.
\end{itemize}
It would be interesting to check quantitatively whether this speculation is borne out in holographic entropy inequalities outside the two infinite families. If so, it would be useful for the `reverse problem': Given a contraction map $f$, find the holographic entropy inequalities that $f$ proves \cite{ningnew}. More generally, we believe that any interpretive framework for a qualitative understanding of holographic entropy inequalities must be intimately related to entanglement wedge nesting.

\acknowledgments
We thank Bowen Chen, Ricardo Esp{\'i}ndola, Keiichiro Furuya, Yichen Feng, Matthew Headrick, Sergio Hern\'andez-Cuenca, Veronika Hubeny, Dimitrios Patramanis, Yunfei Wang, Minjun Xie, Dachen Zhang and Daiming Zhang for discussions. BC thanks the organizers of the Amsterdam String Workshop, the \emph{Quantum Information, Quantum Field Theory and Gravity} program held at ICTS Bangalore, and the conference \emph{Observables in Quantum Gravity: From Theory to Experiment} held at the Aspen Center for Physics, where much of this work was carried out. BC and SS thank the co-organizers and the institutions that supported the workshop \emph{Holographic Duality and Models of Quantum Computation} held at Tsinghua Southeast Asia on Bali, Indonesia, where part of this work was completed. Some figures were previously used in Reference~\cite{Czech:2023xed} of which we are co-authors. This research was supported by an NSFC grant number 12042505, a BJNSF grant under the Gao Cengci Rencai Zizhu program, and a Dushi Zhuanxiang Fellowship.

\appendix
\section{Another projective plane inequality, outside the infinite family}
\label{app:12}

Reference~\cite{Hernandez-Cuenca:2023iqh} announced 1866 holographic entropy inequalities on 7 regions (including the purifier). We characterized the incidence of nesting violations in their proofs by contraction in Section~\ref{sec:375}. In Section~\ref{sec:study} we studied a class of inequalities whose structure is entirely controlled by entanglement wedge nesting, encapsulated in the form of graph models. One may ask whether the 1866 inequalities from \cite{Hernandez-Cuenca:2023iqh} also admit similar graph models and, if so, whether the conclusions of Section~\ref{sec:study} extend there as well. 

Unfortunately, the majority of the 1866 inequalities do not admit useful graph models like the ones studied in Section~\ref{sec:study}. But there are exceptions. We have not determined fully how many of the 1866 inequalities from \cite{Hernandez-Cuenca:2023iqh} admit useful graph models. Instead, to give a flavor of what \emph{can} happen, we analyze one specific inequality from among the 1866.

We focus on the inequality that is listed as number 12 in Reference~\cite{Hernandez-Cuenca:2023iqh}:
\begin{align}
    S_{ABC}+S_{ACD}+S_{ABD}&+S_{ABE}+S_{ACF}+S_{BCEF} \nonumber \\
    &\ge\nonumber \\
    S_{AB}+S_{AC}+S_{AD}+S_{BE}&+S_{CF}+S_{ABCEF}+S_{ABCD}
\end{align}
To highlight its structure and symmetry, we relabel the regions 
$(A)(BCD)(EFO) \rightarrow (O)(A_1 A_2 A_3)(B_1 B_2 B_3)$ and switch some terms for their complements. Eventually, the inequality takes the form:
\begin{align}
 &S_{A_1 A_2 O} + S_{A_2 A_3 O} + S_{A_3 A_1 O}+S_{A_1 B_1 O}+S_{A_2 B_2 O}+S_{A_3 B_3 O} \nonumber \\
 &\qquad\qquad\qquad\qquad\qquad\qquad\ge \nonumber \\
 &S_{A_1 O}+S_{A_2 O}+S_{A_3 O}+S_{A_1 B_1}+S_{A_2 B_2}+S_{A_3 B_3}+S_{B_1 B_2 B_3}
 \label{eq:12th}
\end{align}
This inequality can be represented with a graph. We find that the conclusions of Section~\ref{sec:study} apply in their entirety.

\begin{figure}[t]
    \centering
    \includegraphics[width=0.55\linewidth]{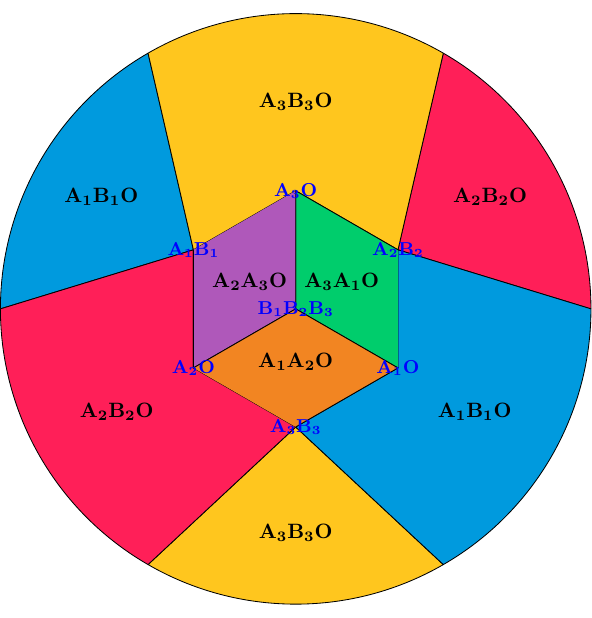}
	\caption{The graph model for inequality~\eqref{eq:12th}. Antipodal points on the boundary of the disk are identified so it is a cross-cap. Thus, the graph is a square tiling of the projective plane.}
         \label{fig:12th}
 \end{figure} 
 
  \begin{figure}[t]
    \centering
    \includegraphics[width=0.9\linewidth]{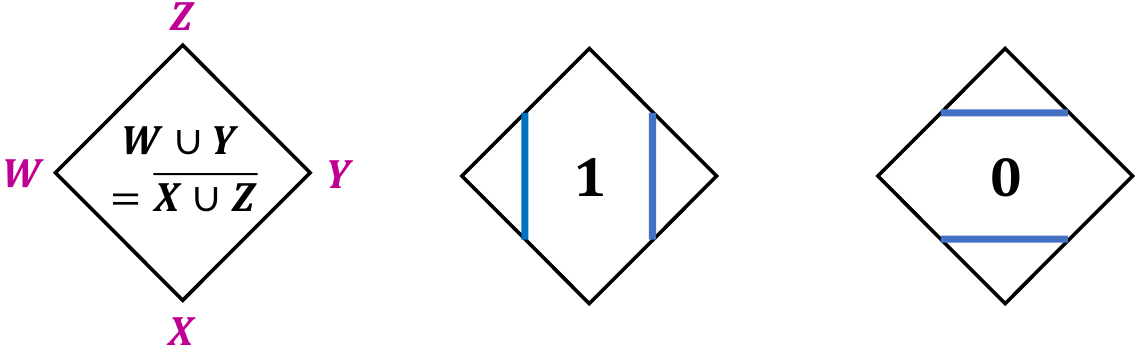}
	\caption{Left panel: A face in Figure~\ref{fig:12th}. Middle and right panels: The auxiliary lines used to define $\Gamma(x)$ in the definition of the contraction map $f$.}
         \label{fig:gamma-app}
 \end{figure} 

\paragraph{Graph} 
We assemble a graph for inequality~(\ref{eq:12th}) following the rules reviewed in Section~\ref{sec:toric}. The result is displayed in Figure~\ref{fig:12th}. We highlight certain salient features of the graph:
 \begin{itemize}
 \item The graph is a tiling of the projective plane. Indeed, it contains $F = 6$ faces, $E = 12$ edges and $V = 7$ vertices so the Euler character is 1. (Note that the boundary circle in Figure~\ref{fig:12th} does not contain edges or vertices; it merely represents a cross-cap.) 
 \item Every face is a quadrilateral. From now on we refer to the faces of Figure~\ref{fig:12th} as `squares.'
 \item Every square has the structure shown in the left panel of Figure~\ref{fig:gamma-app}. That is, if we label the ordered quadruple of vertices around a square as $W, X, Y, Z$ then the square represents $W \cup Y = \overline{X \cup Z} = \overline{X} \cap \overline{Z}$.
 \item One difference with the graphs encountered in Section~\ref{sec:toric} and \ref{sec:rp} is that the edges here do not represent conditional mutual informations.  
 \end{itemize}
  
\paragraph{Proof by contraction}

As in Section~\ref{sec:toricproofreview}, we construct the contraction map as follows:
\begin{itemize}
    \item Set a ranking scheme for the vertices in the graph.
    \item For every bit string $x \in \{0,1\}^6$, we form an auxiliary graph $\Gamma(x)$ according to the rules shown in Figure~\ref{fig:gamma-app}. The graph consists of non-intersecting loops, which divide the projective plane into connected domains. 
    \item The bits of $f(x)$ (labeled by the vertices in Figure~\ref{fig:12th}) are assigned as follows:
\begin{itemize}    
    \item Assign 0 to all vertices in the wrapped connected domain determined by $\Gamma (x)$.
    \item In each unwrapped domain determined by $\Gamma (x)$, assign 1 to the highest-ranked vertex and 0 to all others. 
\end{itemize}
\end{itemize}

Observe that among the connected domains defined by $\Gamma(x)$, always precisely one is wrapped. That is why we use the phrasing `in the wrapped connected domain' above. This observation is confirmed by the following logic: 
\begin{itemize}
\item The fundamental group of the projective plane is $\mathbb{Z}_2$. Thus, the loops of $\Gamma(x)$ are classified as either unwrapped (0) or wrapped (1).
\item A priori, a bit flip on a square can recombine two loops in the following combinations: $0 + 0 \leftrightarrow 0$ or $0 + 1 \leftrightarrow 1$ or $1 + 1 \leftrightarrow 0$.
\item The possibility $1 + 1 \leftrightarrow 0$ above can never occur because two wrapped loops cannot co-exist in $\Gamma(x)$. If they did, they would intersect because the self-intersection number of the non-trivial cycle in $\mathbb{RP}^2$ equals 1.
\item This leaves $0 + 0 \leftrightarrow 0$ and $0 + 1 \leftrightarrow 1$ as the only possible effects that a bit flip can exert on the loops of $\Gamma(x)$. Consequently, if a wrapped loop exists in any one $\Gamma(x)$ then it exists in all of them because no bit flip can ever annihilate it.
\item `Boundary conditions' below show that a wrapped loop does indeed exist in specific $\Gamma(x)$'s.
\end{itemize}

We briefly comment on why this map is a contraction. The logic is identical as in Section~\ref{sec:study} and as highlighted in Figure~\ref{fig:flips}. In short, every bit flip causes one unwrapped loop to be eaten by another loop ($0 + 0 \to 0$ or $0 + 1 \to 1$), or vice versa. This always merges one unwrapped connected domain to another domain, or vice versa. This operation changes the value of $f(x)$ in precisely one bit so that map $f$ is indeed a contraction.

\begin{figure}[t]
		\centering
		$\begin{array}{lp{0.1cm}cp{0.1cm}r}
		\includegraphics[width=0.32\linewidth]{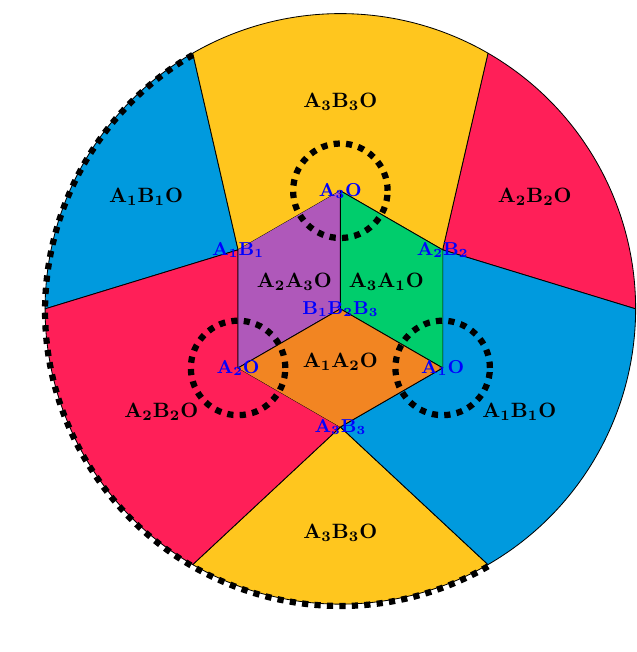} & &
		\includegraphics[width=0.32\linewidth]{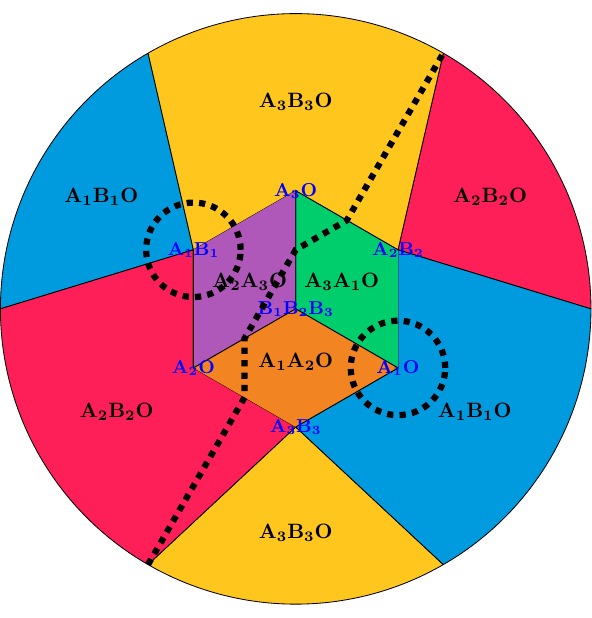} & &
        \includegraphics[width=0.32\linewidth]{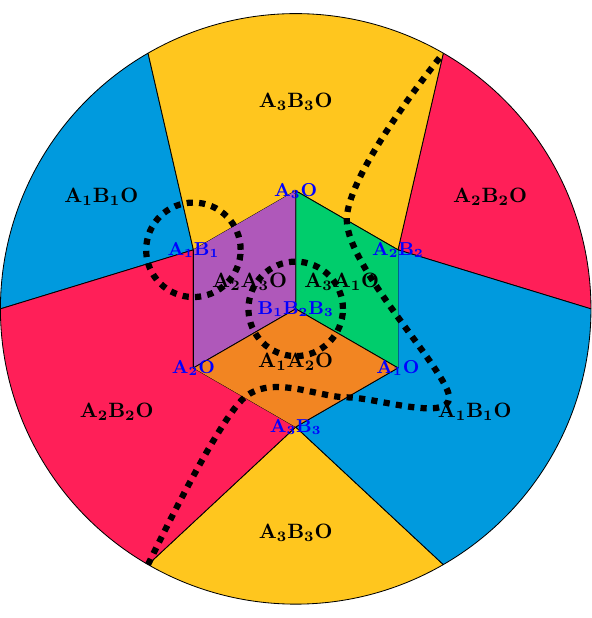}
		\end{array}$
	\caption{Boundary conditions for inequality~\eqref{eq:12th}. The left / middle / right panel displays $\Gamma(x^O)$, $\Gamma(x^{A_1})$, and $\Gamma(x^{B_1})$, respectively. Note the non-contractible loop in $\Gamma(x^O)$, which hugs the cross-cap. The boundary conditions for $A_2, A_3, B_2, B_3$ are obvious $\mathbb{Z}_3$-analogues of the graphs shown here. } 
\label{fig:12th_bc}
\end{figure}

\paragraph{Boundary conditions} and the corresponding auxiliary graphs $\Gamma(x)$ are shown in Figure~\ref{fig:12th_bc}. All unwrapped loops there are minimal. Moreover, they encircle precisely those vertices, which contain the requisite fundamental region. Therefore, $f(x)$ assigns 1 / 0 to vertices that do / do not contain the region in question, and so it respects the required boundary conditions. 

\paragraph{Properties of the contraction map}
We highlight features of the contraction $f$, which mirror the conclusions of Section~\ref{sec:study}:
\begin{itemize}
    \item Map $f$ can assign 1 to a vertex and remain consistent with entanglement wedge nesting only if the vertex is surrounded by a minimal loop in $\Gamma(x)$.  This follows from the way $\Gamma(x)$ is constructed; see Figure~\ref{fig:gamma-app}. 
    \item Map $f$ assigns 1's to vertices greedily, in the sense that if a vertex can be assigned 1 consistent with entanglement wedge nesting, then map $f$ does so.
    \item These assignments of 1's under $f$ are not sufficient to define a contraction. That is, in addition to assigning all 1's that are consistent with entanglement wedge nesting, map $f$ must also assign some 1's that are not consistent with nesting. In other words, nesting violations are necessary to define a contraction.
    \item Unavoidable nesting violations occur whenever two unwrapped connected domains merge into one non-minimal unwrapped domain. By the contraction property, the resulting non-minimal unwrapped domain must contain at least one vertex set to 1. But because the domain is not minimal, assigning the vertices in it to 1 is not consistent with entanglement wedge nesting. In summary, here again non-minimal unwrapped domains are associated with unavoidable nesting violations.
    \item Map $f$ violates nesting only when it is necessary and to the minimal possible extent. That is, it assigns a single 1 per unwrapped connected domain. (If it assigned 1 to more than one vertex in some unwrapped connected domain, it would violate nesting non-minimally. If it assigned 1 to some vertex in the wrapped connected domain, it would be an unnecessary nesting violation.)
    \item At the end of Section~\ref{sec:rp}, we observed that the boundary conditions of the toric and projective plane inequalities are maximally packed with configurations, which are one bit flip away from an unavoidable nesting violation. In terms of the graph $\Gamma(x)$, this looks like a maximally dense arrangement of minimal loops. This ensures that nearly all bit flips---all that act away from the wrapped loop---automatically produce a \emph{non-minimal} unwrapped loop (which always comes with a nesting violation). 
    
In the graph for inequality~(\ref{eq:12th}), it is only possible to build unwrapped loops that encompass 1 vertex (minimal loops) or 2 or 3 vertices (non-minimal loops with unavoidable nesting violations). Because of the small size of the graph, one must take with a grain of salt the claim that here too are boundary conditions arranged for a maximal density of nesting violations. Nevertheless, to tout such a claim, one would observe that in the boundary conditions in Figure~\ref{fig:12th_bc}, the minimal loops are always in maximal proximity to one another. This arrangement indeed maximizes the number of bit flips that can produce non-minimal unwrapped loops and \emph{eo ipso} unavoidable nesting violations.
\end{itemize} 

\bibliographystyle{JHEP}
\bibliography{reference.bib}

\end{document}